\documentclass[useAMS,usenatbib]{mn2e}
\usepackage{graphicx}
\usepackage{amssymb}
\def\lesssim{\!\!\!\phantom{\le}\smash{\buildrel{}\over
 {\lower2.5dd\hbox{$\buildrel{\lower2dd\hbox{$\displaystyle<$}}\over
                               \sim$}}}\,\,}
\usepackage{hyperref}
\usepackage{breakurl}
\newcommand{\urlwofont}[1]{\urlstyle{same}\url{#1}}
\usepackage{upgreek}

\def\mnras{MNRAS}
\def\apj{ApJ}
\def\apjl{ApJ}
\def\aj{AJ}
\def\araa{ARA\&A}
\def\aaps{A\&AS}
\def\aap{A\&A}
\def\pasp{PASP}

\def\procspie{Proc. SPIE}

\def\CaII{Ca\,{\sc ii}} 
 
\def\HeI{He\,{\sc i}}
\def\FeII{Fe\,{\sc ii}}
 
\def\OI{O\,{\sc i}}

\def\MgII{Mg\,{\sc ii}}
\def\NaI{Na\,{\sc i}}
\def\NaID{Na\,{\sc i}~D}

\def\TiII{Ti\,{\sc ii}}
\def\SiII{Si\,{\sc ii}}

\title[SNe 2010O and 2010P in Arp 299 -- I]
{The nature of supernovae 2010O and 2010P in Arp 299 -- I. Near-infrared and optical evolution}
\author[Kankare et al.]
{E. Kankare,$^{1,2}$\thanks{E-mail: eskank@utu.fi} S. Mattila,$^{1}$ S. Ryder,$^{3}$ M. Fraser,$^{4,5}$ A. Pastorello,$^{6}$ N. Elias-Rosa,$^{6,7}$\and 
C. Romero-Ca\~nizales,$^{2,8}$ A. Alberdi,$^{9}$ V.-P. Hentunen,$^{10}$ R. Herrero-Illana,$^{9}$ \and J. Kotilainen,$^{1}$ M.-A. P\'erez-Torres$^{9}$ and P. V\"ais\"anen$^{11,12}$ \\
$^{1}$ Finnish Centre for Astronomy with ESO (FINCA), University of Turku, V\"ais\"al\"antie 20, FI-21500 Piikki\"o, Finland\\
$^{2}$ Tuorla Observatory, Department of Physics and Astronomy, University of Turku, V\"ais\"al\"antie 20, FI-21500 Piikki\"o, Finland\\
$^{3}$ Australian Astronomical Observatory, PO Box 915, North Ryde, NSW 1670, Australia\\
$^{4}$ Institute of Astronomy, University of Cambridge, Madingley Road, Cambridge CB3 0HA, UK\\
$^{5}$ School of Mathematics and Physics, Queens University Belfast, Belfast BT7 1NN, UK\\
$^{6}$ INAF -- Osservatorio Astronomico di Padova, Vicolo dell'Osservatorio 5, I-35122 Padova, Italy\\
$^{7}$ Institut de Ci\`encies de l'Espai (IEEC-CSIC), Facultat de Ci\`encies, Campus UAB, E-08193 Bellaterra, Spain\\
$^{8}$ Instituto de Astrof\'{\i}sica, Facultad de F\'{\i}sica, Pontificia Universidad Cat\'olica de Chile, Casilla 306, Santiago 22, Chile \\
$^{9}$ Instituto de Astrof\'isica de Andaluc\'ia, IAA-CSIC, Apartado 3004, E-18080 Granada, Spain\\
$^{10}$ Taurus Hill Observatory, H\"ark\"am\"aentie 88, FI-79480 Kangaslampi, Finland\\
$^{11}$ South African Astronomical Observatory, PO Box 9, Observatory 7935, Cape Town, South Africa\\
$^{12}$ Southern African Large Telescope, PO Box 9, Observatory 7935, Cape Town, South Africa}

\date{Accepted 2013 November 22.
      Received 2013 November 22;
      in original form 2013 October 13}
\begin{document}

\maketitle

\label{firstpage}

\begin{abstract}
We present near-infrared and optical photometry, plus optical spectroscopy of two stripped-envelope supernovae (SNe) 2010O and 2010P that exploded in two different components of an interacting luminous infrared galaxy Arp~299 within only a few days of one another. SN~2010O is found to be photometrically and spectroscopically similar to many normal Type Ib SNe and our multiwavelength observations of SN~2010P suggest it to be a Type IIb SN. No signs of clear hydrogen features or interaction with the circumstellar medium are evident in the optical spectrum of SN~2010P. We derive estimates for the host galaxy line-of-sight extinctions for both SNe, based on both light-curve and spectroscopic comparison finding consistent results. These methods are also found to provide much more robust estimates of the SN host galaxy reddening than the commonly used empirical relations between extinction and equivalent width of \NaID\ absorption features. The SN observations also suggest that different extinction laws are present in different components of Arp~299. For completeness, we study high-resolution pre-explosion images of Arp~299 and find both SNe to be close to, but not coincident with, extended sources that are likely massive clusters. A very simple model applied to the bolometric light curve of SN~2010O implies a rough estimate for the explosion parameters of $E_{\mathrm{k}} \approx  3 \times 10^{51}$~erg, $M_{\mathrm{ej}} \approx 2.9$~M$_{\sun}$ and $M_{\mathrm{Ni}} \approx  0.16$~M$_{\sun}$. 

\end{abstract}


\section{Introduction}

Luminous and ultraluminous infrared galaxies (LIRGs and ULIRGs) are actively star-forming galaxies with high infrared (IR) luminosity ($10^{11} < L_{{\rm IR}} < 10^{12} L_{\sun}$ and $L_{{\rm IR}} > 10^{12} L_{\sun}$, respectively). The Arp~299 system of multiple galaxy components is commonly adopted as a prototypical LIRG due to its IR luminosity ($L_{\mathrm{IR}} = L[$8$-$1000$~\upmu \mathrm{m}] \approx 6.7 \times 10^{11} L_{\sun}$; \citealt{sanders03}), relatively nearby distance ($\sim$44.8~Mpc; \citealt{huo04}) and interacting nature. Arp~299 consists of two main components, Arp~299A and Arp~299B (sometimes referred to as IC~694 and NGC~3690, respectively), both likely to host an active galactic nucleus \citep[see e.g.][]{dellaceca02, perez-torres10}. Most of the star formation in LIRGs is concentrated in their central regions, which is also the case in Arp~299. The milliarcsecond resolution of very long baseline interferometry (VLBI) radio observations has proven that there is a rich population of radio supernovae (SNe) and young SN remnants in the nuclear regions of Arp~299A and Arp~299B, within only $\sim$100$-$150 and $\sim$30~pc from the centre, respectively \citep{perez-torres09,ulvestad09,romero11,bondi12}. In particular, \citet{bondi12} reported a lower limit of core-collapse supernova (CCSN) rate of $\sim$0.80~yr$^{-1}$ for Arp~299A, suggesting that the bulk of the current star formation was taking place in the innermost $\sim$150~pc. The most recent estimate for the total CCSN rate of Arp~299 is 1.6$-$1.9~yr$^{-1}$ derived by \citet{mattila12} based on IR and radio observations. So far, Arp~299 has hosted seven optically detected SNe and has been continuously the target of multiple systematic SN search programmes in the optical and near-IR (NIR) wavelengths during the last $\sim$15 yr \citep[for a summary see][]{mattila12}. 

Recently \citet*{anderson11} carried out a study on the seven optically detected SNe in Arp~299 indicating that the stripped-envelope Type Ib and IIb SNe are more common and more centrally concentrated than other CCSNe in Arp~299 compared to normal spiral galaxies in the local Universe. The authors explain the excess of Type Ib and IIb SNe to be caused either by the young age of the dominating star formation or a top-heavy initial mass function (IMF), see also \citet*{habergham10}. In the former case of still-young starburst only the most massive stars would have reached the end of their life cycle, exploding as stripped-envelope SNe, in contrast with Type II SNe with progenitors having lower mass and longer lifetimes. In the latter model the skewed IMF produces an intrinsically higher ratio of high mass progenitors. However, these are both predicated on the assumption that the progenitors to stripped-envelope SNe are massive, presumably Wolf Rayet like stars, whereas the results may reflect a more complex progenitor population of stripped-envelope SNe. In fact, a number of recent studies \citep[e.g.][]{dessart12, eldridge13} have concluded that binaries must contribute a significant fraction of stripped-envelope SN progenitors. The reason why binary systems could be more common in LIRGs such as Arp~299 is unclear, although this can be related to a larger fraction of star formation taking place in high-density environments of young massive clusters \citep*{portegieszwart10}. A similar trend of excess stripped-envelope SNe has also recently been found in statistical samples of disturbed \citep*{habergham12} and IR bright actively star-forming galaxies \citep{kangas13}, who argued metallicity effects not to be a dominant effect on this. We also note that the seemingly `normal' spiral galaxy NGC~2770 has hosted three CCSNe which have all been Type Ib SNe \citep{thone09} showing that such a CCSN distribution can also exist in a non-starburst galaxy by chance.  

This is Paper~I in a set of two papers studying the nature of SNe~2010O and 2010P, the two most recent optically detected SNe in Arp~299, the initial results of which were used by the recent studies of \citet{anderson11} and \citet{mattila12}. In this paper we present and discuss in detail the NIR and optical follow-up data of these SNe, including adaptive optics (AO) data obtained with the Gemini-North Telescope. The accompanying Paper~II \citep[][hereafter Paper~II]{romero14} concentrates on the radio data obtained on SN~2010P probing the SN interaction with the surrounding circumstellar medium (CSM). 

In Section~2 background information on SNe~2010O and 2010P is given, in Section~3 we present the NIR and optical observations and in Section~4 the results from the analysis are reported. The high-resolution pre-explosion data are studied in Section~5 and the nature of the two SNe is discussed in Section~6. Conclusions are given in Section~7.

\section{SN 2010O and SN 2010P}

\begin{figure*}
\includegraphics[width=\linewidth]{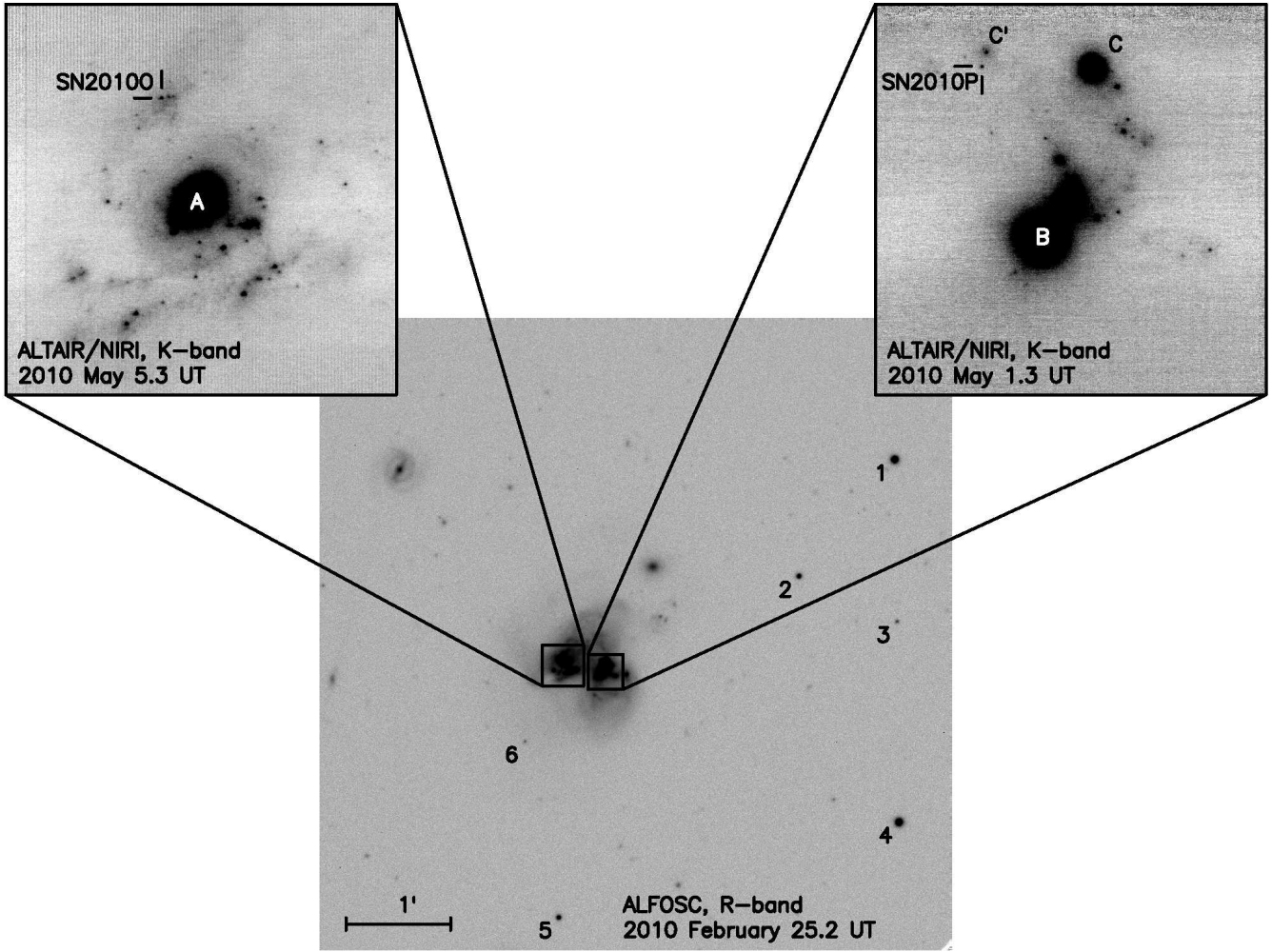}
\caption{$6 \times 6$~arcmin$^{2}$ \textit{R}-band subsection of ALFOSC image of the Arp~299 field, and $20 \times 20$~arcsec$^{2}$ subsections of \textit{K}-band ALTAIR/NIRI AO images of SN~2010O and SN~2010P. Sequence stars used to calibrate the non-AO photometry are marked in the \textit{R}-band image. IR bright components of Arp~299 are marked in the \textit{K}-band images following the notation of \citet{gehrz83}. North is up and east is to the left in the images.}
\label{fig:Arp299}
\end{figure*}

The discovery of SN~2010O, located at $\alpha = 11^{\mathrm{h}} 28^{\mathrm{m}} 33\fs86$ and $\delta = +58\degr 33\arcmin 51\farcs6$ (equinox J2000.0), was first reported by \citet*{newton10} based on unfiltered optical imaging carried out by the Puckett Observatory Supernova Search on 2010 January 24.37 and 25.34 {\sc ut}. Arp~299 was also observed by us in a monthly monitoring programme on 2010 January 18.2 {\sc ut} using the 2.56-m Nordic Optical Telescope (NOT) in \textit{JHKs} bands, leading to a NIR discovery of both SNe~2010O and 2010P via comparison to a reference image, with SN~2010P located at $\alpha = 11^{\mathrm{h}} 28^{\mathrm{m}} 31\fs38$ and $\delta = +58\degr 33\arcmin 49\farcs3$ (equinox J2000.0). A confirming epoch of observations was obtained on 2010 January 23.1 {\sc ut}. Thus our independent NIR discovery of SN~2010O precedes that of \citet{newton10} and provides tighter constraints on the explosion date. The fact that the Puckett Observatory did not report the discovery of SN~2010P already indicates a very high visual host galaxy extinction in the line-of-sight to this SN. Our NIR discovery of SN~2010P was reported in \citet{mattila10}. The field of SNe~2010O and 2010P is shown in Fig.~\ref{fig:Arp299}, with an \textit{R}-band image of Arp~299 and subsections of the \textit{K}-band Altitude Conjugate Adaptive Optics for the Infrared (ALTAIR)/Near-Infrared Imager (NIRI) AO images of the main components of Arp~299. The explosion site of SN~2010O is overlapping with one of the spiral arms of the A component of Arp~299, whereas the explosion site of SN~2010P has a small angular distance (170~pc) to the IR bright C$\arcmin$ nucleus of Arp~299. For the naming convention of the components of Arp~299, see \citet*{gehrz83}.

\citet{beswick10} reported a 5.0-GHz Multi-Element Radio Linked Interferometer Network (MERLIN) non-detection of SNe~2010O and 2010P at the 3$\sigma$ confidence level of 186~$\upmu$Jy~beam$^{-1}$, observed between 2010 January 29.8 and February 1.1 {\sc ut}. The first radio detection of SN~2010P was reported by \citet{herrero-illana12}, obtained on 2011 June 15, $\sim$1.4~yr from the explosion, with the Expanded Very Large Array (EVLA) at 8.5~GHz with a 703 $\pm$ 60~$\upmu$Jy~beam$^{-1}$ counterpart to SN~2010P. 

The optical spectroscopy of SN~2010O obtained on 2010 January 28.0 {\sc ut} showed it to be a Type Ib SN close to maximum light \citep{mattila10b}. The radio non-detections of SN~2010O, reported in Paper~II, are consistent with this \citep{soderberg06a}. An optical spectrum of SN~2010P was obtained on 2010 February 11.5 {\sc ut} showing it to be consistent with a very highly reddened Type Ib or Type IIb SN, 1$-$3 weeks after the maximum light, as reported in \citet{ryder10}. However, the Type Ib origin of SN~2010P is ruled out by the late-time evolution observed at radio wavelengths, indicating that the SN is a Type IIb event, with a substantially longer rise time to reach the peak radio luminosity than previously observed for any Type IIb SN (see Paper~II). 

\citet{bond10} obtained early ground-based images of SN~2010O and by aligning them with pre-explosion archival \textit{Hubble Space Telescope} (\textit{HST}) images observed in \textit{B} and \textit{I} bands (\textit{F435W} and \textit{F814W}, respectively) reported the SN to be coincident with a young and very blue stellar cluster hosting likely very massive stars. However, the follow-up observations with the \textit{HST} revealed SN~2010O to be only close, but not on the line-of-sight to this cluster \citep{bond11}. Furthermore, \citet{nelemans10} analysed two epochs of \textit{Chandra} X-ray archival data concluding the position of SN~2010O to be coincident with a varying X-ray source and suggested its progenitor to be a Wolf-Rayet X-ray binary, if associated with this system.

\section{Observations}

\subsection{Near-infrared and optical imaging}

\textit{JHKs}-band imaging of Arp~299 with the NOT was carried out with the Nordic Optical Telescope Near-Infrared Camera and Spectrograph \citep[NOTCam;][]{djupvik10} with a pixel scale of 0.234~arcsec~pixel$^{-1}$ and field-of-view (FOV) of $4 \times 4$~arcmin$^{2}$ in the wide-field imaging mode. The NOTCam data reduction included flat-fielding, sky subtraction and median combination of the aligned frames. The reductions were carried out using an external {\sc iraf}\footnote{{\sc iraf} is distributed by the National Optical Astronomy Observatories, which are operated by the Association of Universities for Research in Astronomy, Inc., under cooperative agreement with the National Science Foundation.} package {\sc notcam} version 2.5, available on the NOT webpages.\footnote{\urlwofont{http://www.not.iac.es/instruments/notcam/guide/observe.html}} The geometric distortion on the NOTCam was also corrected using the above mentioned package. \textit{JHKs} images obtained on 2009 July 15.0 {\sc ut} in our monitoring programme were used as reference frames for image subtraction. Reference frames were reduced similarly to the frames with the SNe, and the image pairs were aligned with {\sc iraf} scripts based on {\sc geomap} and {\sc geotran}. Image subtraction was carried out with the {\sc isis}~2.2 image subtraction package based on the Optimal Image Subtraction (OIS) method \citep{alard98, alard00}. In the OIS method the point spread function (PSF) of the better seeing image is matched to that of the worse seeing image by using a convolution kernel derived by the software between the spatially aligned pair of images. The NOTCam images of SNe~2010O and 2010P were calibrated using Two Micron All-Sky Survey (2MASS) stars in the FOV. The PSF photometry was carried out with the {\sc quba}\footnote{{\sc python} package specifically designed by S. Valenti for SN imaging and spectra reduction. For more details on the pipeline, see \citet{valenti11}.} pipeline.

Follow-up imaging of SNe~2010O and 2010P with the Gemini-North Telescope was carried out as part of our survey to search for highly obscured nuclear CCSNe in a sample of eight LIRGs \citep[see][]{kankare08, kankare12}. The observations were conducted with the NIRI \citep{hodapp03} with the ALTAIR laser guide star (LGS) AO system in \textit{JHK}-bands (programmes GN-2009B-Q-23 and GN-2010A-Q-40, PI: S. Ryder). The ALTAIR/NIRI AO set-up with the f32 camera provides a pixel scale of 0.0219~arcsec~pixel$^{-1}$, a FOV of $22.4 \times 22.4$~arcsec$^{2}$ and a typical full width at half-maximum (FWHM) of 0.1~arcsec. The ALTAIR/NIRI data were reduced with the external {\sc niri} package available in {\sc iraf}, including flat-fielding and sky subtraction. We used our own {\sc iraf} scripts to remove the horizontal noise pattern visible in some of the frames and to combine the individual frames. Similar to the NOTCam data, reference images were subtracted from the reduced ALTAIR/NIRI frames using {\sc isis}~2.2. The reference frames were obtained with the same instrument and set-up in \textit{JHK} on 2012 January 29.5 {\sc ut} for SN~2010O, in \textit{K} on 2011 April 17.3 {\sc ut} and in \textit{JH} on 2012 January 31.5 {\sc ut} for SN~2010P. The subtracted images used the zero-point derived from the reference fields, and were calibrated with instrument zero-points derived from photometric \textit{JHK}-band standard star \citep{hawarden01} observations. The standard star FS~127 was observed on the same nights as the reference frames with the exception of the \textit{K}-band reference field for SN~2010P, which was calibrated with the photometric standard star FS~131 observed on the following night. The atmospheric extinction correction was done using the values of \citet{leggett06}. The photometry was carried out using aperture photometry with {\sc gaia}.\footnote{\urlwofont{http://star-www.dur.ac.uk/~pdraper/gaia/gaia.html}} For each subtracted image a curve of growth of the SN flux with increasing aperture size was measured and an optimal aperture size was selected to be as large as possible such that the aperture photometry would not be compromised by nearby sources or background, with a typical radius adopted of 0.7 or 0.9~arcsec. This was used to measure both the SN and to derive the instrument zero-point. The statistical error given by {\sc gaia} was adopted as the error of photometry, no systematic errors were estimated. This is likely to somewhat underestimate the errors of ALTAIR/NIRI photometry, when the SNe were still bright. 

The optical imaging of SNe~2010O and 2010P was obtained using the Andalucia Faint Object Spectrograph and Camera\footnote{The data presented here were obtained in part with ALFOSC, which is provided by the Instituto de Astrofisica de Andalucia (IAA) under a joint agreement with the University of Copenhagen and NOTSA.} (ALFOSC) and stand-by CCD camera (StanCam) at the NOT, the RATCam at the Liverpool Telescope \citep[LT;][]{steele04}, and the SBIG ST-8XME camera with the 14-inch Celestron at Taurus Hill Observatory (THO). Archive images from the Wide Field Camera (WFC) at the Isaac Newton Telescope (INT) and the \textit{HST} Wide Field Camera 3 (WFC3) Ultraviolet-Visible (UVIS) channel were also used. The ground-based imaging was reduced using the standard {\sc iraf} tasks and the help of the {\sc quba} pipeline, including bias subtraction and flat-field correction. Similar to the NIR data, the optical images were template subtracted using the OIS method, and the reference images of Arp~299 obtained with the ALFOSC on 2009 April 13.1 {\sc ut} and 2012 February 18.2 {\sc ut}. The optical images were calibrated using (up to) six field stars which, in turn, were calibrated with the observations of the field of Arp~299 and \citet{landolt92} standard star fields (SA~104~334 and PG~1323$-$086) on 2013 May 28.1 {\sc ut}. The night on La Palma was later confirmed as photometric based on the atmospheric extinction monitoring measurements at the Carlsberg Meridian Telescope\footnote{\urlwofont{http://www.ast.cam.ac.uk/~dwe/SRF/camc_extinction.html}} and the comparison of the derived and the reported\footnote{\urlwofont{http://www.not.iac.es/instruments/alfosc/zpmon/}} instrumental zero-points for ALFOSC to each other. The PSF-fitting photometry was carried out with the {\sc quba} pipeline. The optical colour terms for ALFOSC and RATCam were adopted from \citet{ergon13}. The image subtraction yielded detections of SN~2010O in the \textit{BVRI} bands whereas SN~2010P could only be detected in the \textit{I} band due to the high line-of-sight host galaxy extinction obscuring the SN in shorter wavebands. The magnitudes of the sequence stars derived in \textit{BVRI} and adopted from 2MASS in \textit{JHK} are listed in Table~\ref{table:sequence} and the photometry for SNe~2010O and 2010P is presented in Tables~\ref{table:phot_10O_nir}$-$\ref{table:phot_10P}. In 2010 July SNe~2010O and 2010P disappeared behind the Sun. However, when Arp~299 became observable again at a reasonable airmass, \textit{JHKs} imaging with NOTCam was obtained on 2010 November 1.3 {\sc ut}, which yielded non-detection of both SNe. 

PSF-fitting photometry was performed on both SNe in the archive {\it HST}+WFC3/UVIS images (programme GO-12295, PI: H.~Bond) taken on 2010 June 24.9 {\sc ut} using the {\sc dolphot} package.\footnote{\urlwofont{http://americano.dolphinsim.com/dolphot/}} The magnitudes are reported in the VEGAMAG system in the \textit{HST} filters, which are comparable to \textit{UBI} for \textit{F336W}, \textit{F438W} and \textit{F814W}, respectively.

\begin{table*}
\caption{Magnitudes of the Arp~299 field stars (for the identifications, see Fig.~\ref{fig:Arp299}). The errors are given in brackets.}
\centering
\begin{tabular}{cccccccc}
\hline
Star & $m_{B}$ & $m_{V}$ & $m_{R}$ & $m_{I}$ & $m_{J}$ & $m_{H}$ & $m_{K}$ \\
\# & (mag) & (mag) & (mag) & (mag) & (mag) & (mag) & (mag) \\ 
\hline
1 & 15.49(0.01)	& 14.91(0.01) & 14.56(0.02) & 14.21(0.01) & 13.82(0.03) & 13.55(0.04) & 13.52(0.05)\\
2 & 18.26(0.01)	& 17.21(0.01) & 16.46(0.02) & 15.82(0.01) & 14.94(0.05) & 14.50(0.07) & 14.27(0.07)\\
3 & 20.97(0.01) & 19.41(0.01) & 18.44(0.02) & 17.10(0.01) & 15.83(0.09) & 15.38(0.14) & 15.03(0.12)\\
4 & 15.78(0.01) & 14.85(0.01) & 14.37(0.02) & 13.87(0.01) & 13.22(0.02) & 12.79(0.03) & 12.67(0.03)\\
5 & 18.21(0.01) & 17.13(0.01) & 16.87(0.02) & 16.24(0.01) & 15.56(0.07) & 14.84(0.09) & 15.36(0.16)\\
6 & 21.16(0.01) & 19.87(0.01) & 19.27(0.02) & 18.39(0.01) & $-$ & $-$ & $-$ \\
\hline
\end{tabular}
\label{table:sequence}
\end{table*}

\begin{table*}
\caption{NIR photometry for SN~2010O. The errors are given in brackets.}
\centering
\begin{tabular}{cccccc}
\hline
JD & Date & $m_{J}$ & $m_{H}$ & $m_{K}$ & Telescope/instrument\\
(240\,0000+) & ({\sc ut}) & (mag) & (mag) & (mag) & \\ 
\hline
55214.7 & January 18.2 & 16.55(0.03) & 16.49(0.05) & 16.22(0.07) & NOT/NOTCam\\
55219.5 & January 23.0 & 16.19(0.02) & 16.06(0.05) & 15.64(0.05) & NOT/NOTCam\\
55224.1 & January 27.6 & 15.93(0.04) & 15.32(0.02)  & 14.74(0.02) & Gemini-North/ALTAIR/NIRI\\
55234.8 & February 7.3 & 15.92(0.02) & 14.94(0.09) & 14.85(0.13) & NOT/NOTCam\\
55251.9 & February 24.4 & $-$  & 15.77(0.02) & 15.45(0.05) & Gemini-North/ALTAIR/NIRI\\
55252.9 & February 25.4 & 16.95(0.10) & $-$  & $-$ & Gemini-North/ALTAIR/NIRI\\
55315.5 & April 29.0 & 18.41(0.12) & 17.68(0.13) & 17.44(0.12) & NOT/NOTCam\\
55321.8 & May 5.3 & 18.83(0.42) & 17.21(0.09) & 17.29(0.20) & Gemini-North/ALTAIR/NIRI\\
\hline
\end{tabular}
\label{table:phot_10O_nir}
\end{table*}

\begin{table*}
\caption{Optical photometry for SN~2010O. The errors are given in brackets.}
\centering
\begin{tabular}{cccccccc}
\hline
JD & Date & $m_{U}$ & $m_{B}$ & $m_{V}$ & $m_{R}$ & $m_{I}$ & Telescope/instrument\\
(240\,0000+) & ({\sc ut}) & (mag) & (mag) & (mag) & (mag) & (mag) & \\ 
\hline
55219.6 & January 23.1 & $-$ & $-$ & $-$ & $-$ & 16.55(0.03) & NOT/StanCam\\
55223.4 & January 26.9 & $-$ & 18.27(0.11) & 17.41(0.02) & $-$ & 16.45(0.04) & THO\\
55224.5 & January 28.0 & $-$ & 18.46(0.05) & 17.48(0.02) & 16.99(0.03) & 16.43(0.02) & NOT/ALFOSC\\
55237.5 & February 10.0 & $-$ & 19.63(0.05) & 18.02(0.08) & 17.30(0.03) & 16.54(0.02) & LT/RATCam\\
55240.5 & February 13.0 & $-$ & 19.79(0.07) & 18.20(0.05) & 17.44(0.03) & 16.68(0.06) & LT/RATCam\\
55252.7 & February 25.2 & $-$ & 20.61(0.15) & 18.98(0.05) & 18.06(0.03) & 17.30(0.02) & NOT/ALFOSC \\
55264.5 & March 9.0 & $-$ & 21.04(0.15) & 19.20(0.06) & 18.42(0.02) & 17.61(0.03) & NOT/ALFOSC \\
55275.4 & March 19.9 & $-$ & $-$ & $-$ & 18.70(0.09) & 17.64(0.10) & NOT/ALFOSC\\
55345.4 & May 28.9 & $-$ & 21.68(0.26) & 20.60(0.10) & 19.47(0.14) & 19.10(0.15) & INT/WFC\\
55372.4 & June 24.9 & 22.90(0.04) & 22.03(0.02) & $-$ & $-$ & 19.77(0.01) & \textit{HST}/WFC3\\  
\hline
\end{tabular}
\label{table:phot_10O_opt}
\end{table*}

\begin{table*}
\caption{Photometry for SN~2010P. The errors are given in brackets.}
\centering
\begin{tabular}{ccccccc}
\hline
JD & Date & $m_{I}$ & $m_{J}$ & $m_{H}$ & $m_{K}$ & Telescope/instrument\\
(240\,0000+) & ({\sc ut}) & (mag) & (mag) & (mag) & (mag) & \\ 
\hline
55214.7 & January 18.2 & $-$ & 17.47(0.04) & 16.84(0.07) & 16.34(0.06) & NOT/NOTCam\\
55219.5 & January 23.0 & $-$ & 17.04(0.04) & 16.48(0.04) & 15.77(0.05) & NOT/NOTCam\\
55219.6 & January 23.1 & 19.36(0.03) & $-$ & $-$ & $-$ & NOT/StanCam\\
55224.1 & January 27.6 & $-$ & $-$ & 16.06(0.03) & 15.25(0.03) & Gemini-North/ALTAIR/NIRI\\
55224.5 & January 28.0 & 19.08(0.05) & $-$ & $-$ & $-$ & NOT/ALFOSC\\
55234.8 & February 7.3 & $-$ & 16.30(0.03) & 15.07(0.11) & 15.23(0.28) & NOT/NOTCam\\ 
55237.5 & February 10.0 & 19.11(0.17) & $-$ & $-$ & $-$ & LT/RATCam\\
55240.5 & February 13.0 & 19.13(0.09) & $-$ & $-$ & $-$ & LT/RATCam\\
55252.7 & February 25.2 & 19.45(0.24) & $-$ & $-$ & $-$ & NOT/ALFOSC\\
55252.9 & February 25.4 & $-$ & 17.22(0.10) & 16.17(0.03) & 15.37(0.03) & Gemini-North/ALTAIR/NIRI\\
55264.5 & March 9.0 & 19.75(0.10) & $-$ & $-$ & $-$ & NOT/ALFOSC\\
55315.5 & April 29.0 & $-$ & 19.06(0.13) & 17.81(0.06) & 17.13(0.08) & NOT/NOTCam\\
55317.8 & May 1.3 & $-$ & $-$ & 17.68(0.11) & 17.20(0.12) & Gemini-North/ALTAIR/NIRI\\
55372.4 & June 24.9 & 22.18(0.02) & $-$ & $-$ & $-$ & \textit{HST}/WFC3\\
\hline
\end{tabular}
\label{table:phot_10P}
\end{table*}

\subsection{Optical spectroscopy}

An optical spectrum of SN~2010O was obtained at the NOT with the ALFOSC on 2010 January 28.0 {\sc ut}. Also a low-resolution optical spectrum of SN~2010P was obtained during the same night with ALFOSC, however, due to the high extinction it had a very low signal-to-noise ratio. A follow-up spectrum of SN~2010O was also obtained on 2010 March 20.1 {\sc ut} with ALFOSC. The NOT spectra were reduced using the {\sc quba} pipeline including bias subtraction, flat-fielding, spectral extraction, wavelength and flux calibration. The wavelength calibration was carried out using a HeNe arc lamp exposure and cross-correlated with the sky lines. Relative flux calibration was conducted using a spectrum of a spectroscopic standard star observed close in time to the SN spectrum. Telluric features have not been removed. 

A deep optical spectrum of SN~2010P was obtained with the Gemini-North Telescope using the Gemini Multi-Object Spectrograph \citep[GMOS;][]{hook04} on 2010 February 11.5 {\sc ut}. The GMOS optical spectral data were reduced using the {\sc gmos} tasks in the {\sc gemini} package within {\sc iraf}. A master bias frame (constructed by averaging with 3$\sigma$ clipping a series of bias frames) was subtracted from all raw images. Images of a CuAr lamp spectrum were used to wavelength calibrate the science images and straighten them along the spatial dimension, while images of a quartz-halogen lamp spectrum helped correct for sensitivity variations within and between the CCDs. Observations of the spectrophotometric standard star Feige 66 \citep{massey88} with the same instrument set-up and similar airmass to that of the SN~2010P observations enabled a system response function to be derived and applied over the wavelength range 4710$-$8950~\AA; and the removal of telluric features in the SN 2010P spectrum. The spectroscopic observations are reported in Table~\ref{table:spect_log}.

\begin{table*}
\begin{center}
\caption{Log of spectroscopy}
\begin{tabular}{cccccccc}
\hline
SN & JD & Date & Grism/grating & Range & Resolution & Exp. time & Telescope/instrument\\ 
& (240\,0000+) & ({\sc ut}) & & (\AA) & (\textit{R}) & (s) & \\ \hline
2010O & 55224.5 & January 28.0 & gm\#7 & 3850$-$6850 & 650 & 600 & NOT/ALFOSC \\
2010P & 55239.0 & February 11.5 & R400 & 4000$-$9500 & 1918 & $4 \times 900$ & Gemini-North/GMOS \\
2010O & 55275.6 & March 20.1 & gm\#4 & 3200$-$9100 & 270 & 3600 & NOT/ALFOSC\\ 
\hline
\end{tabular}
\label{table:spect_log}
\end{center}
\end{table*}

\section{Results}

\subsection{Photometric results}
\label{ssec:nir}

\begin{table*}
\caption{Parameters of the SNe adopted for the photometric and spectroscopic comparison with SNe~2010O and 2010P.}
\centering
\begin{tabular}{ccccccccccc}
\hline
SN & JD$_{\mathrm{explosion}}$ & $\mu$ & $A_{V}^{\mathrm{Gal}}$ & $A_{V}^{\mathrm{host}}$ & JD$_{\mathrm{max}}$ & $t_{\mathrm{max}}$ & $M_{I}^{\mathrm{max}}$ & $M_{i'}^{\mathrm{max}}$ & $M_{J}^{\mathrm{max}}$ & Ref. \\
 & (240\,0000+) & (mag) & (mag) & (mag) & (240\,0000+) & (d) & (mag) & (mag) & (mag) & \\ 
\hline
1999ex & 51480.5 & 33.55 & 0.06 & 0.87 & 51498.1$^{a}$ & 18$^{a}$ & $-18.0$ & $-$ & $-$ & 1\\  
2005bf & 53458 & 34.62 & 0.14 & 0.0$^{b}$ & 53499.8$^{c}$ & 42$^{c}$ & $-$ & $-18.4$ & $-$ & 2,3 \\
2007Y & 54145.5 & 31.43 & 0.07 & 0.28 & 54163.1$^{c}$ & 18$^{c}$ & $-$ & $-16.3$ & $-16.7$ & 4 \\
2009jf & 55099.5 & 32.65 & 0.35 & 0.16 & 55120.0$^{c}$ & 20.5$^{c}$ & $-$ & $-$ & $-$ & 5 \\ 
2010O & 55203.7 & 33.26 & 0.05 & $\sim$2 & 55225$^{c}$ & 19$^{c}$ & $-17.8$ & $-$ & $-17.9$ & 6,7\\
\vspace{-0.1in}\\
1993J & 49074.0 & 27.71 & 0.22 & 0.36 & 49094.5$^{d}$ & 20.5$^{d}$ & $-17.6$ & $-$ & $-17.9$ & 7,8,9\\
2000H & - & 33.66 & 0.63 & 0.0 & 51586$^{e}$ & $-$ & $-$ & $-$ & $-$ & 7,10,11\\
2008ax & 54528.8 & 29.92 & 0.07 & 1.17 & 54547.1$^{e}$ & 18.3$^{e}$ & $-17.75$ & $-$ & $-17.80$ & 12\\
2010P & 55206.7 & 33.26 & 0.05 & $\sim$7 & $\sim$55228$^{c}$ & $\sim$21$^{c}$ & $-17.9$ & $-$ & $-19.2$ & 6,7\\
2011dh & 55713.0 & 29.46 & 0.09 & 0.12 & $\sim$55733.8$^{c}$ & 20.8$^{c}$ & $-17.42$ & $-$ & $-17.58$ & 13\\
\hline
\end{tabular}
\begin{flushleft}
$^{a}$ $L_{B}$ maximum.\\
$^{b}$ $E(B-V) \approx 0.1$~mag also possible according to \citet{folatelli06}.\\
$^{c}$ $L_{bol}$ maximum.\\
$^{d}$ Second $L_{B,V}$ maximum.\\ 
$^{e}$ $L_{V}$ maximum.\\ 
References: 1 $-$ \citet{stritzinger02}; 2 $-$ \citet{tominaga05}; 3 $-$ \citet{folatelli06}; 4 $-$ \citet{stritzinger09};\\ 
5 $-$ \citet{valenti11}; 6 $-$ This work; 7 $-$ \citet{schlafly11}; 8 $-$ \citet{lewis94}; 9 $-$ \citet{mattila01};\\ 
10 $-$ \citet{branch02}; 11 $-$ \citet{elmhamdi06}; 12 $-$ \citet{taubenberger11}; 13 $-$ \citet{ergon13}.
\end{flushleft}
\label{table:parameters}
\end{table*}

The optical+NIR light curves were used to estimate the host galaxy line-of-sight extinction and evaluate the best-fitting extinction law towards SNe~2010O and 2010P. The \textit{BVRIJH} light curves of SN~2010O and \textit{IJHK} light curves of 2010P were compared to those of the Type Ib SN~2007Y \citep{stritzinger09} and Type IIb SN~2011dh \citep{ergon13}, respectively. Light curves or templates of other CCSN types were not tried for the comparison, as the classification of the SNe based on optical spectroscopy and radio observations was considered to be robust. SN~2007Y was selected as the comparison template for SN~2010O as they are spectroscopically similar (see Section~\ref{ssec:optical_10O}) and there are well sampled broad-band light curves available, covering SN~2007Y also around maximum light, including the crucial NIR region. The general transformations of \citet{smith02} were used to convert the \textit{r}$\arcmin$- and \textit{i}$\arcmin$-band magnitudes of SN~2007Y \citep{stritzinger09} into \citet{landolt92} system \textit{R}- and \textit{I}-band magnitudes for the comparison. SN~2011dh was selected as the comparison template for SN~2010P, as it has (i) the best sampled optical and NIR light curves available in the literature for any Type IIb event; and (ii) like SN~2010P, it does not show either a double-peaked shape in the light curves, nor strong H$\alpha$ features in the post-maximum spectrum (see Section~\ref{ssec:optical_10P}) unlike the prototypical Type IIb SN~1993J. 

The light-curve comparisons were carried out as a $\chi^{2}$-fitting between the observed and the template light curves similar to \citet{kankare12}. Three free parameters were estimated, i.e. the epoch $t_{0}$ of the first detection from the explosion, \textit{V}-band host galaxy extinction $A_{V}$ and a constant shift $C$ applied to the comparison light curves. All the light curves used in the comparison, in different wavebands, were fitted simultaneously to match those of the chosen template SN to derive the optimal set of parameters taking into account all the photometric data in a single fit to minimize the $\chi^{2}$ value. We emphasize that the same derived constant $C$ is applied to all bands of the template light curve, reflecting the difference in the absolute brightness between the SNe and possible errors in distances. The intrinsic colours of the comparison and reference SNe are assumed to be the same. As the parameter $C$ does not affect the colour of the compared SNe, it has negligible effect on the derived values for host galaxy extinction. Both $A_{V}$ and $C$ are tightly related to the observed colour of the SN and a minimum number of wavebands required for a reliable estimate of these parameters is three. By contrast, the epoch $t_{0}$ is defined mainly by the evolution of the light curves and to estimate the epoch of the first detection requires data over a long time frame. In this study we have multiple epochs of data, both pre- and post-maximum, covering SN~2010O in six and SN~2010P in four different bands over a broad wavelength range. Therefore, no degeneracy issues are present in the fitting process.

For Arp~299 we adopt a Galactic extinction of $A_{V} = 0.05$~mag \citep{schlafly11} and a luminosity distance of 44.8~Mpc corrected to the cosmic microwave background reference frame \citep{fixsen96} corresponding to a distance modulus of $\mu = 33.26 \pm 0.15$ ($H_0 = 73$\,km\,s$^{-1}$\,Mpc$^{-1}$). Both the \citet*[][$R_{V}=3.1$]{cardelli89} and \citet[][$R_{V}=4.05$]{calzetti00} extinction laws were considered for the host galaxy extinction of SNe~2010O and 2010P. The \citet{calzetti00} extinction, or attenuation, curve is strictly speaking not applicable to a single point source, but since it is widely used to describe the dust properties of starburst galaxies in particular, and in the absence of a well-defined extinction law for LIRGs, we use it in comparison to the \citet{cardelli89} extinction law. For the Galactic extinction and host galaxy extinction of the comparison SNe, the \citet{cardelli89} extinction law was adopted. For a summary of the parameters of SNe used to compare with SN~2010O and SN~2010P photometrically and spectroscopically, see Table~\ref{table:parameters}. 

\begin{figure*}
\includegraphics[width=0.7\linewidth]{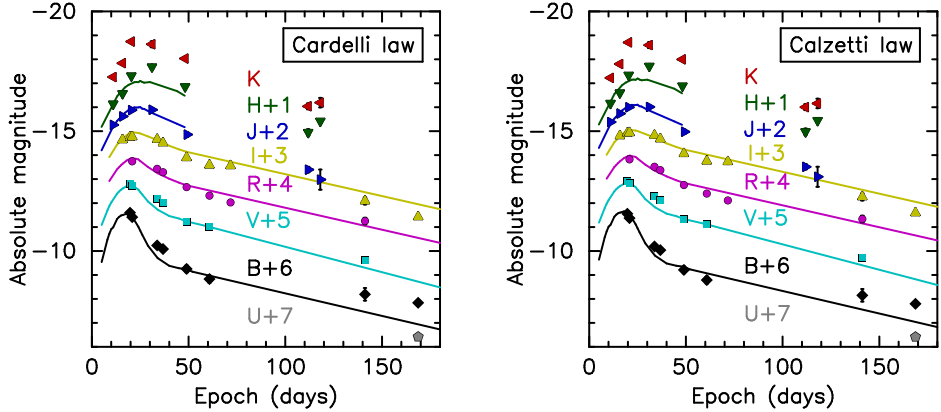}
\caption{Light curves of SN~2010O (points) compared with those of the Type Ib SN~2007Y \citep{stritzinger09} in \textit{BVRIJH} (lines) via  $\chi^{2}$ fitting. The SN absolute magnitudes have been corrected for the derived total line-of-sight extinctions adopting the Cardelli extinction law on the left and the Calzetti extinction law on the right. The template light curves have been shifted in magnitude relative to SN~2010O by a constant $C$. The epoch is relative to the explosion date derived as part of the fitting. The fitting parameters are reported in Table~\ref{table:chi2}.}
\label{fig:lc_10O}
\end{figure*}

\begin{figure*}
\includegraphics[width=0.7\linewidth]{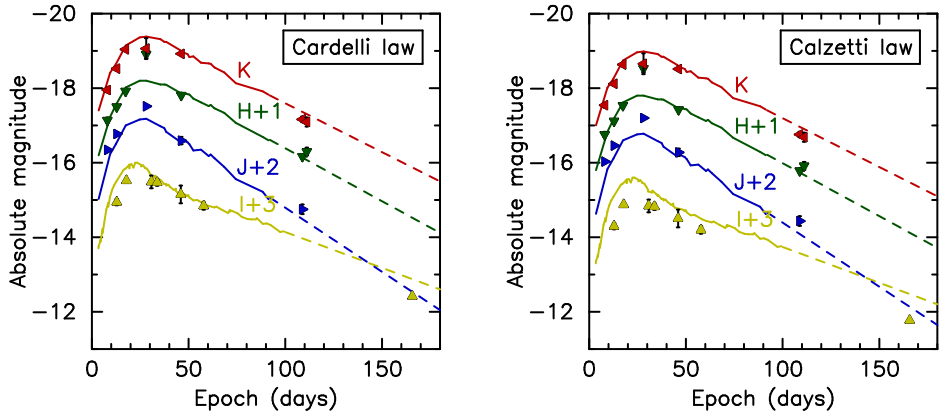}
\caption{As Fig.~\ref{fig:lc_10O} but for SN~2010P compared with the Type IIb SN~2011dh \citep{ergon13} using $\chi^{2}$ fitting in \textit{IJHK} bands. The template light curves have been linearly extrapolated for illustrative purposes.}
\label{fig:lc_10P}
\end{figure*}

The $\chi^{2}$-fitting provides reasonable matches between SN~2010O and SN~2007Y with both host galaxy extinction laws and suggests our NIR discovery of SN~2010O to have been roughly 11 d after the explosion (2010 January 7), SN~2010O to be $\sim$1.4~mag brighter than the low-luminosity Type Ib SN~2007Y and to be extinguished by $A_{V}\approx2$~mag of host galaxy extinction. The template comparison for SN~2010P suggests it to be discovered roughly 8 d after the explosion (2010 January 10), to be $\sim$1.4~mag brighter than SN~2011dh and to have a line-of-sight host galaxy extinction of $A_{V}=$~7$-$9~mag, depending on the adopted extinction law. The $\chi^{2}$-fit values are reported in Table~\ref{table:chi2} and the absolute magnitude light curves of SNe~2010O and 2010P are shown in Figs~\ref{fig:lc_10O} and \ref{fig:lc_10P} including also the template light curves (shifted with a constant $C$), that provided the best fit. We also note that the non-detections of SNe~2010O and 2010P in our ALTAIR/NIRI images of Arp~299 from 2009 December 3 and 5, respectively, are consistent with a relatively early discovery of the two SNe, though not providing strong constraints. 

When adopting for SN~2010O the best $\chi^{2}$-fit using the \citet{cardelli89} extinction law with $A_{V}=1.9$~mag, a line-of-sight extinction corrected colour (\textit{V}$-$\textit{R})$ = 0.24 \pm 0.09$~mag is obtained roughly 14 d from the assumed \textit{V}-band peak. This is in excellent agreement with the intrinsic colour of (\textit{V}$-$\textit{R})$ = 0.26 \pm 0.06$~mag for Type Ib/c SNe on day 10 post \textit{V}-band maximum, derived empirically by \citet{drout11}. 

\begin{table}
\caption{Best $\chi^{2}$ fits for the parameters for SNe~2010O and 2010P.}
\centering
\begin{tabular}{ccccc}
\hline
Extinction law & $A_{V}$ & $t_{0}$ & $C$ & $\tilde{\chi}^{2}$\\
 & (mag) & (d) & (mag) & \\ 
\hline
SN 2010O & & & & \\
\vspace{-0.1in}\\
Cardelli law & 1.9 & 11 & $-1.3$ & 33.3 \\
Calzetti law & 2.0 & 11 & $-1.4$ & 25.2 \\
\vspace{-0.1in}\\
SN 2010P & & & & \\
\vspace{-0.1in}\\
Cardelli law & 9.0 & 8 & $-1.6$ & 23.4 \\
Calzetti law & 6.8 & 8 & $-1.2$ & 40.6 \\
\hline
\end{tabular}
\label{table:chi2}
\end{table}

\subsection{Bolometric light curve of SN~2010O}

An interpolated pseudo-bolometric \textit{UBVRIJHK} light curve of SN~2010O was created from available photometry assuming host galaxy extinction of $A_{V}=1.9$~mag and the \citet{cardelli89} extinction law. The \textit{U}-band contribution of SN~2010O was estimated by upscaling the same bandpass magnitudes of SN~2007Y, based on the \textit{BVRIJH} $\chi^{2}$-comparison, as detailed in the previous section. The \textit{U}-band magnitudes of SN~2007Y were converted from the \textit{Swift} Ultraviolet/Optical Telescope observations reported by \citet{brown09} using the method presented by \citet{li06}. The possible errors arising from this necessary approach of estimating the \textit{U}-band contribution from a comparison SN can be assumed to be relatively small, as for Type Ib/c SNe, the total contribution of the UV is only $\sim$5$-$15~per~cent of the total flux \citep{valenti08}. To derive the pseudo-bolometric light curve, fluxes from the extinction-corrected magnitudes were integrated over the filter functions and converted into luminosities. 

\begin{figure*}
\includegraphics[width=0.7\linewidth]{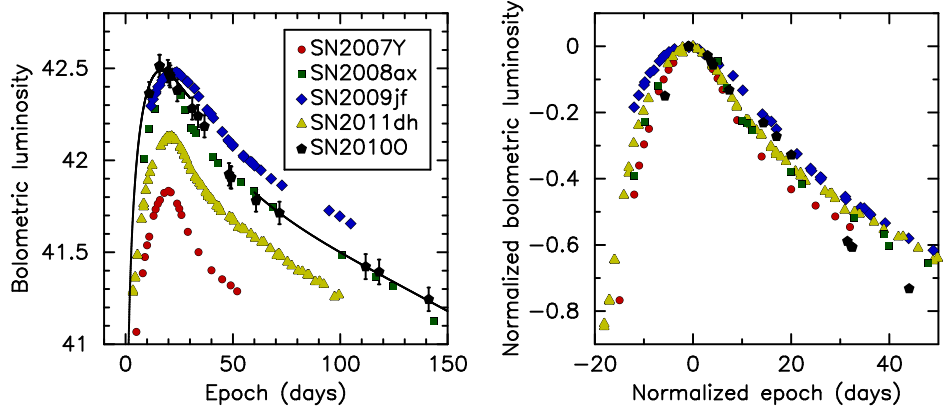}
\caption{On the left the pseudo-bolometric \textit{UBVRIJHK} light curve of SN~2010O is compared to those of Type Ib SNe~2007Y \citep{stritzinger09} and 2009jf \citep{valenti11}, and Type IIb SNe~2008ax \citep{taubenberger11} and 2011dh \citep{ergon13}. In the same plot the best fit from the two-phase model of \citet{valenti08} is shown. On the right the comparison is shown, normalized with the epoch and luminosity of the bolometric light-curve peak.}
\label{fig:lc_bol}
\end{figure*}

To infer first-order estimates of the explosion parameters, kinetic energy $E_{\mathrm{k}}$, ejecta mass $M_{\mathrm{ej}}$ and $^{56}$Ni mass $M_{\mathrm{Ni}}$ of SN~2010O we fitted to the bolometric light curve a simple model presented by \citet{valenti08} based on the work of \citet{arnett82}. For a full and detailed description of the model and its initial assumptions and assumed parameters, we refer to \citet{valenti08} and references therein, and provide only a short summary on the main features of the model here. The model fit for SN~2010O was carried out in two parts: one covering the photospheric phase of the SN (the initial 30 d), and a second one covering its nebular phase (from 60 d since the explosion and onwards). The fitting of the photospheric phase is based on the model of \citet{arnett82} appropriate for Type I SNe, expanded by \citet{valenti08} to include not only the radioactive decay of $^{56}$Ni but also decay of $^{56}$Co. During the nebular phase the energy output of the SN is dominated by the radioactive decay of $^{56}$Co. In the work of \citet{valenti08} this energy output, taking into account the incomplete $\gamma$-ray and e$^{+}$ trapping, is described by equations presented in \citet{sutherland84}, \citet{cappellaro97} and \citet{clocchiatti97}. The model takes into account the energy arising from $\gamma$-rays from $^{56}$Ni and $^{56}$Co decay and the annihilation of the e$^{+}$ created in the cobalt decay, as well as the kinetic energy of these positrons \citep{valenti08}. The fitting to the bolometric light curve of SN~2010O was carried out including a two-component model of inner high-density and high-velocity ejecta and low-density outer ejecta following the photospheric bulk velocity at maximum luminosity, adopted from the \FeII\ $\lambda$5169 line velocity, see Section~\ref{ssec:optical_10O}. In Fig.~\ref{fig:lc_bol} the best model fit is shown together with a comparison of the pseudo-bolometric \textit{UBVRIJHK} light curve of SN~2010O with a selection of other stripped-envelope SNe. The constant optical opacity of 0.06~cm$^{2}$~g$^{-1}$ was adopted in the fitting \citep{valenti11}. The method yields a rough estimate of  $E_{\mathrm{k}}(\mathrm{total}) \approx  3 \times 10^{51}$~erg, $M_{\mathrm{ej}}(\mathrm{total}) \approx 2.9$~M$_{\sun}$ and $M_{\mathrm{Ni}}(\mathrm{total}) \approx  0.16$~M$_{\sun}$, with an inner ejecta contribution of $E_{\mathrm{k}}(\mathrm{inner}) \approx 0.01 \times 10^{51}$~erg, $M_{\mathrm{ej}}(\mathrm{inner}) \approx 0.65$~M$_{\sun}$ and $M_{\mathrm{Ni}}(\mathrm{inner}) \approx 0.04$~M$_{\sun}$. The comparison in Fig.~\ref{fig:lc_bol} shows the bolometric light curve of SN~2010O to be fairly similar to that of SN~2008ax for which \citet{valenti11} inferred relatively similar explosion parameters.  

\subsection{Spectroscopic results for SN~2010O}
\label{ssec:optical_10O}

\begin{figure}
\includegraphics[width=\linewidth]{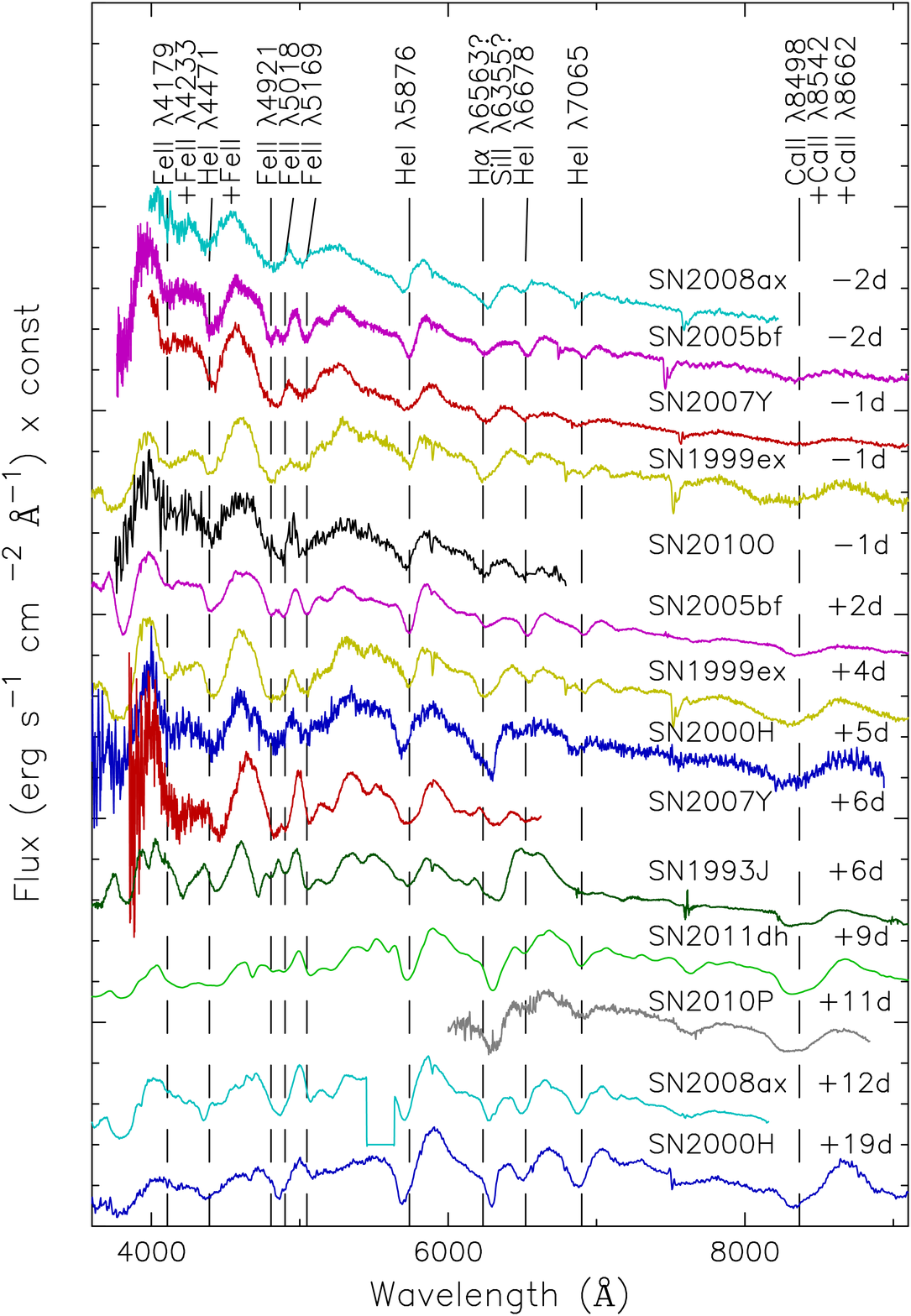}
\caption{The classification spectra of SNe~2010O and 2010P placed in a sequence of spectroscopically normal and similar Type Ib and IIb SNe: SN~1999ex \citep{hamuy02}, SN~2000H \citep{branch02}, SN~2005bf \citep{folatelli06}, SN~2007Y \citep{stritzinger09}, SN~2008ax \citep{pastorello08} and SN~2011dh \citep{ergon13}. Spectrum of the extended Type IIb SN~1993J \citep{barbon95} included as a comparison to SN~2010P on a similar epoch shows a more prominent H$\alpha$ feature. The spectra are dereddened and the wavelengths corrected to the host galaxy rest frame with a host galaxy extinction of $A_{V}=1.9$~mag and \citet{cardelli89} extinction law adopted for SN~2010O and $A_{V}=6.8$~mag and \citet{calzetti00} extinction law adopted for SN~2010P, as suggested by the light-curve comparison. The spectra are also multiplied and vertically shifted with arbitrary constants for clarity. Main spectroscopic features in the spectra, blueshifted by 7000~km~s$^{-1}$ to match roughly the absorption minima of the SNe, are marked. The comparison spectra have been obtained from the Weizmann interactive supernova data repository \citep{yaron12}.}
\label{fig:spect_10O}
\end{figure}

The optical classification spectrum of SN~2010O is similar to spectroscopically normal Type Ib events, such as SNe~1999ex \citep{hamuy02}, 2005bf \citep{folatelli06} or 2007Y \citep{stritzinger09}, close to the maximum light. We do note that SN~2005bf was a peculiar SN with a slowly evolving double-peak optical light curve \citep{tominaga05} and a slow rise to the maximum bolometric luminosity as late as day 42 from the inferred explosion date, which has been associated with a very asymmetric explosion \citep{folatelli06}. In the literature the spectra of SN~2005bf close in time to the first of the two light-curve peaks have been compared to normal Type Ib SNe close to maximum light \citep[e.g.][]{folatelli06}. However, we suggest to place the classification spectrum of SN~2010O in a sequence of spectra around maximum luminosity of the Type Ib SNe~1999ex \citep{hamuy02} and 2007Y \citep{stritzinger09}, as well as SN~2005bf, see Fig.~\ref{fig:spect_10O}. The spectrum comparison shows a coherent sequence of evolution supporting the epoch of the spectrum of SN~2010O to be close in time to the maximum bolometric luminosity. This is also quite consistent with the light-curve comparison with SN~2007Y, which suggested our first NIR detection to have taken place 11 d after the explosion. This would make the epoch of our optical spectrum to be 21 d from the explosion. For comparison, SN~2007Y peaked in luminosity roughly at day 18. 

The dominant features in the classification spectrum of SN~2010O are helium and iron lines, in particular \HeI\ $\lambda\lambda$4471, 5876, 6678 and \FeII\ $\lambda\lambda$4921, 5018, 5169 \citep[see e.g.][]{hamuy02,anupama05}. In addition the feature at $\sim$4100~\AA\ is likely a blend of \FeII\ $\lambda\lambda$4179, 4233 and the \HeI\ $\lambda$4471 is possibly blended with multiple \FeII\ lines. Other elements such as \MgII\ and \TiII\ may also be blended in some of the observed line features. We adopt the \FeII\ $\lambda$5169 line as the best tracer of the photospheric velocity, for which a Gaussian fit to the absorption minimum suggests a photospheric velocity $v_{\mathrm{ph}} \sim$~9000~km~s$^{-1}$ near the light-curve peak. \HeI\ lines in the spectrum of SN~2010O show somewhat lower velocities. In addition, the spectrum shows an absorption feature at $\sim$6240~\AA\, which has also been observed in some other stripped-envelope SNe, spectroscopically also similar to SN~2010O, and has been associated e.g. with \SiII\ $\lambda\lambda$6347, 6371, high-velocity H$\alpha$, a blend of these features or some other metals or their blends \citep[e.g.][and references therein]{folatelli06, stritzinger09}. In the case of SN~2005bf the feature was considered to arise, at least partly, from high-velocity, $\sim$15000~km~s$^{-1}$, H$\alpha$ due to the similar high velocity of the feature associated with H$\beta$ and lines clearly identified as \CaII\ \citep[see e.g. the discussions of][]{folatelli06, parrent07, stritzinger09}. If associated with \SiII\ the line velocity of the feature would be roughly 6000~km~s$^{-1}$ and thus slower than the inferred photospheric velocity, though we note that this is not impossible. As noted above, the spectrum of SN~2010O is very similar to that of SN~2005bf close to the peak. If the feature in SN~2010O is associated with high-velocity H$\alpha$, at $\sim$15000~km~s$^{-1}$, this would indicate that at least a small amount of hydrogen was still present in the outer envelope of the progenitor. Unfortunately our spectrum does not cover the \CaII\ NIR triplet for further comparison of the velocities.

\begin{figure}
\includegraphics[width=\linewidth]{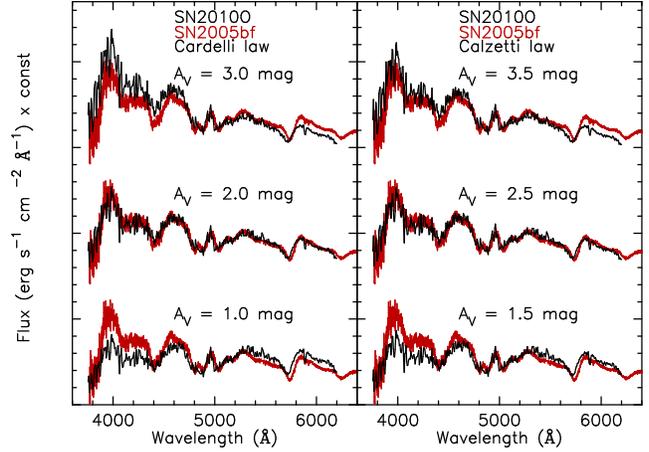}
\caption{Selection of dereddened spectra of SN~2010O using both the \citet{cardelli89} or \citet{calzetti00} extinction law and compared to the dereddened spectrum of SN~2005bf obtained 2 d before the maximum bolometric luminosity \citep{folatelli06}.}
\label{fig:spect_10O_05bf}
\end{figure}

Because of the striking similarity between the classification spectrum of SN~2010O and the spectra of SN~2005bf at the time of the maximum luminosity a more detailed comparison was carried out with different applied host galaxy extinctions and two extinction laws, those of \citet{cardelli89} and \citet{calzetti00}. The spectrum of SN~2010O, shifted in the rest wavelength, was dereddened with extinction in the range of A$_{V} = $~0$-$10~mag in steps of 0.1~mag. The dereddened spectra were scaled to the spectrum of SN~2005bf \citep[dereddened with the extinction law of][]{cardelli89} and subtracted from this reference spectrum. The standard deviation of the subtraction was used as a measure of the quality of the fit and the minimum value was assumed to infer the optimal host galaxy extinction yielding $A_{V} \approx 2.0$ or $2.5$~mag assuming the \citet{cardelli89} or \citet{calzetti00} extinction laws, respectively. The best comparison and cases with $\pm$1~mag of extinction for reference are shown in Fig.~\ref{fig:spect_10O_05bf}. Part of the spectrum, redwards from 6200~\AA, has been omitted from the comparison due to the second-order effect, caused by the ALFOSC grism \#7, which was used to obtain the classification spectrum of SN~2010O. We conclude that the \textit{BVRIJH} light-curve comparison and the optical spectrum comparison are both consistent and favour the line-of-sight host galaxy extinction for SN~2010O to be $A_{V} \approx 2$~mag and to follow the \citet{cardelli89} extinction law. We note that the spectra of SNe~2010O and 2010P (Section~\ref{ssec:optical_10P}) are compared to reference SNe with quite small estimated host galaxy line-of-sight extinctions ($\lesssim0.1$~mag). Any errors that these extinction estimates might have are additive to the inferred results. Compared to the high line-of-sight extinctions of the two SNe in Arp~299 that are studied here, these errors are expected to be very small.

\begin{figure}
\includegraphics[width=\linewidth]{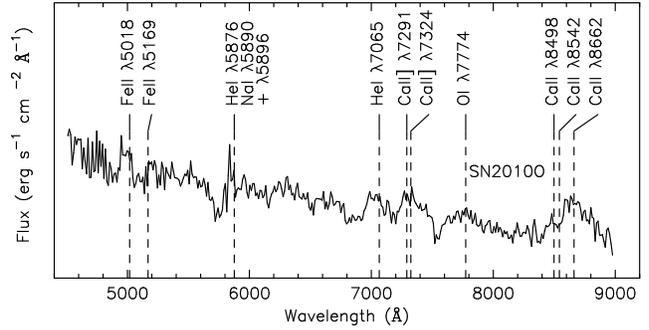}
\caption{The optical spectrum of SN~2010O roughly 50 d from the maximum light. Similar to Fig.~\ref{fig:spect_10O} the spectrum is dereddened and the wavelengths corrected to the host galaxy rest frame.}
\label{fig:spect_10O_2}
\end{figure}

The second spectrum of SN~2010O was obtained 51 d after the classification spectrum. The extended wavelength range reveals also the detection of the \HeI\ $\lambda$7065 feature not covered by the first spectrum. \HeI\ $\lambda$5876 is also visible, possibly blended with \NaI\ $\lambda\lambda$5890, 5896 doublet. The spectrum also reveals \CaII\ NIR triplet, the \OI\ $\lambda$7774 feature and the semiforbidden \CaII] $\lambda\lambda$7291, 7324 doublet. The spectrum still shows some continuum; however, the onset of the latter features is indicating that the SN is moving towards the nebular phase. Similarly, very tentatively the forbidden [\OI] $\lambda\lambda$6300, 6364 can be detected in SN~2010O, arising from the ejecta synthesized in the SN explosion. The spectrum is shown in Fig.~\ref{fig:spect_10O_2} with the most prominent features identified. 

\subsection{Spectroscopic results for SN~2010P}
\label{ssec:optical_10P}

The spectrum of SN~2010P was already shown in \citet{mattila12} as a part of an investigation of the missing fraction of SNe in LIRGs. In Section~\ref{ssec:nir} the discovery date of SN~2010P was inferred to be only 8 d after the explosion when comparing the light curves of SN~2010P to those of SN~2011dh. Based on this, the epoch of the spectrum corresponds to roughly 12 d after the bolometric maximum of SN~2011dh, quite consistent with the spectral evolution of several Type Ib/IIb SNe, as was shown in Fig.~\ref{fig:spect_10O}. Part of the SN~2010P spectrum, bluewards from 6000~\AA, was omitted from the analysis due to unreliable relative flux calibration. The spectrum shows a \CaII\ NIR triplet blend, \HeI\ $\lambda$7065 line and a weak H$\alpha$ feature, similar to such Type IIb SNe as 2000H, 2008ax and 2011dh. 

\begin{figure}
\includegraphics[width=\linewidth]{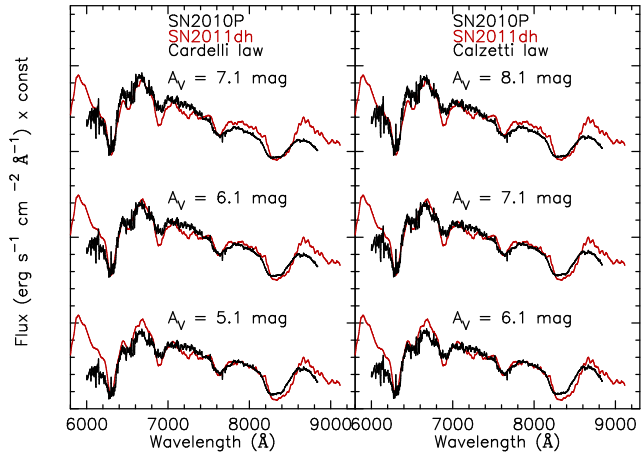}
\caption{Dereddened spectra of SN~2010P, assuming either \citet{cardelli89} or \citet{calzetti00} extinction laws, are compared to the dereddened spectrum of SN~2011dh \citep{ergon13} at a similar epoch.}
\label{fig:spect_10P_11dh}
\end{figure}

Similar to the analysis of SN~2010O, a comparison between the dereddened spectra of SN~2010P with the reference spectrum of SN~2011dh at a similar epoch was carried out. The method yielded an estimate of $A_{V} \approx 6.1$ or $7.1$~mag for the host galaxy extinction of SN~2010P assuming the \citet{cardelli89} or \citet{calzetti00} extinction laws, respectively. The comparison is shown in Fig.~\ref{fig:spect_10P_11dh}. Between the light-curve and spectrum comparisons of SN~2010P there is a clear discrepancy in the derived extinctions if the \citet{cardelli89} extinction law is adopted. However, if the \citet{calzetti00} extinction law is adopted, even though it provides a worse fit in the \textit{I} band, the derived results from the light-curve and the spectrum comparisons are in fact in pretty good agreement, even if the methods and wavelength regions probed are quite different. To conclude, we favour a host galaxy extinction of roughly 7~mag in \textit{V} band for SN~2010P and that the line-of-sight extinction law is following that of \citet{calzetti00}. The derived extinction is also consistent, though slightly higher, than the initial estimate of \citet{mattila12} and does not affect their results.

\subsection{\NaID\ features}
\label{ssec:NaID}

A commonly used method of estimating the host galaxy line-of-sight extinction to SNe is to measure the equivalent width (EW) of the \NaI\ $\lambda\lambda$5890, 5896 doublet and apply an empirical relation inferred between extinction and the \NaID\ EW. However, recently \citet{poznanski11} argued, based on a sample of hundreds of low-resolution Type Ia SN spectra, that no strong correlation exists between the extinction and the EW measured for \NaID. Nonetheless, the method is widely used to give an estimate for CCSNe in the absence of a superior method. To carry out a comparison to our host galaxy extinction estimates, Gaussian components (with the same FWHM) were fitted to both the \NaID$_{1}$ and D$_{2}$ absorption features in the spectra of SNe~2010O and 2010P, using the {\sc spectool} task in {\sc iraf}. This yielded EW(D$_{1}) \approx 1.4$~\AA\ and EW(D$_{2}) \approx 1.1$~\AA\ for SN~20100 and EW(D$_{1}) \approx 2.4$~\AA\ and EW(D$_{2}) \approx 3.0$~\AA\ for SN~2010P. Unreliable relative flux calibration of SN~2010P bluewards from 6000~\AA\ does not prevent measuring \NaID\ EW as it is normalized with the continuum. Following for example the commonly used relations of \citet*{turatto03} with the \citet{cardelli89} extinction law, this suggests a host galaxy extinction of $A_{V}=1.2$~mag (or $A_{V}=3.8$~mag with the steeper relation) and $A_{V}=2.6$~mag (or $A_{V}=8.4$~mag) for SN~2010O and SN~2010P, respectively. Both estimates for both SNe are in discrepancy with our previously derived values for the host galaxy line-of-sight extinction. 

\begin{figure}
\includegraphics[width=\linewidth]{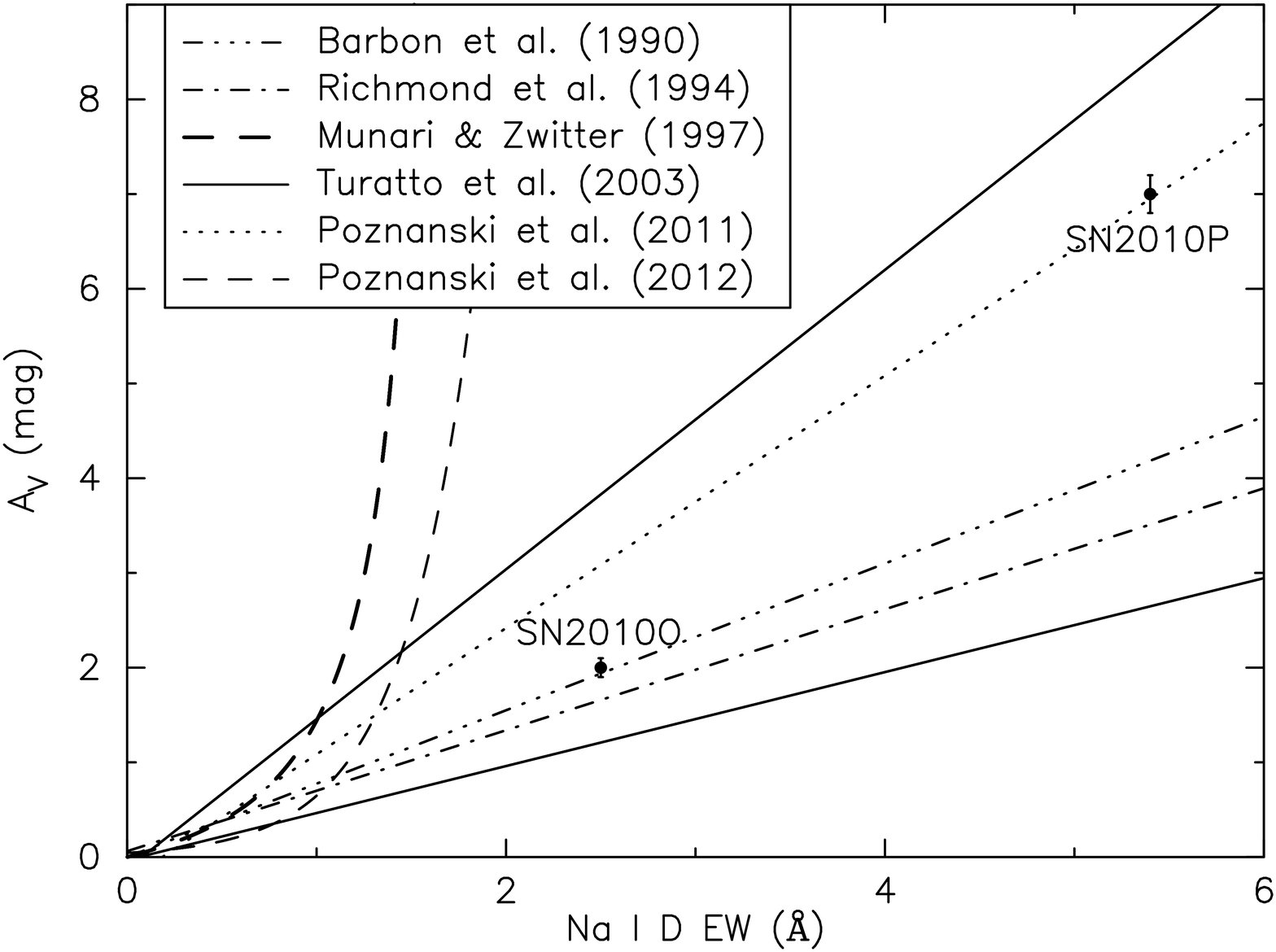}
\caption{\NaID\ EW and $A_{V}$ values of SNe~2010O and 2010P shown together with a selection of empirical relations by \citet{barbon90}, \citet{richmond94}, \citet{munari97}, \citet{turatto03} and \citet{poznanski11, poznanski12}.}
\label{fig:plot_NaID}
\end{figure}

A selection of empirical \NaID\ EW versus $A_{V}$ relations is plotted in Fig.~\ref{fig:plot_NaID} together with our derived results for SNe~2010O and 2010P. The relations of \citet{barbon90}, \citet{turatto03} and \citet{poznanski11} are based on SN data, whereas \citet{richmond94} and \citet{munari97} make use of stellar spectra, and \citet*{poznanski12} use the spectra of galaxies. In particular, the logarithmic relations would suggest unrealistically high extinctions for SNe~2010O and 2010P based on the \NaID\ EW measurements. However, none of the empirical relations is unambiguously consistent with both SNe, at least when using low-resolution spectra. This is likely reflecting the large intrinsic scatter in such relations and can be related to the non-negligible saturation effect of the \NaID\ lines \citep{sollerman05}. Overall this suggests that, especially with high host galaxy extinctions, broad wavelength range light-curve and spectroscopic comparisons, as carried out here, are likely to provide much more robust estimates of the reddening than the methods based on the usage of \NaID\ EW. In particular, high-resolution NIR observations have already been shown to be crucial when investigating host galaxy extinctions for SNe that explode in highly obscured ($A_{V}\gtrsim10$~mag) environments \citetext{\citealp[SN~2008cs,][]{kankare08}; \citealp[SN~2008iz,][]{mattila13}}.

\section{Pre-explosion imaging}

Whilst there have been numerous successful identifications of the progenitors of CCSNe in archival images \citep[see][for a review]{smartt09}, this has typically only been accomplished for very nearby SNe ($\lesssim 25$ Mpc). The distance of Arp~299 (44.8~Mpc) and the high extinction towards SNe~2010O and 2010P ($A_{V} = 2$ and 7~mag, respectively) make the identification of a progenitor candidate in this instance unlikely. Nonetheless, we discuss below the available pre-explosion archival data taken with the \textit{HST}.

\subsection{SN 2010O}

\begin{figure*}
\centering
  \includegraphics[width=0.32\linewidth]{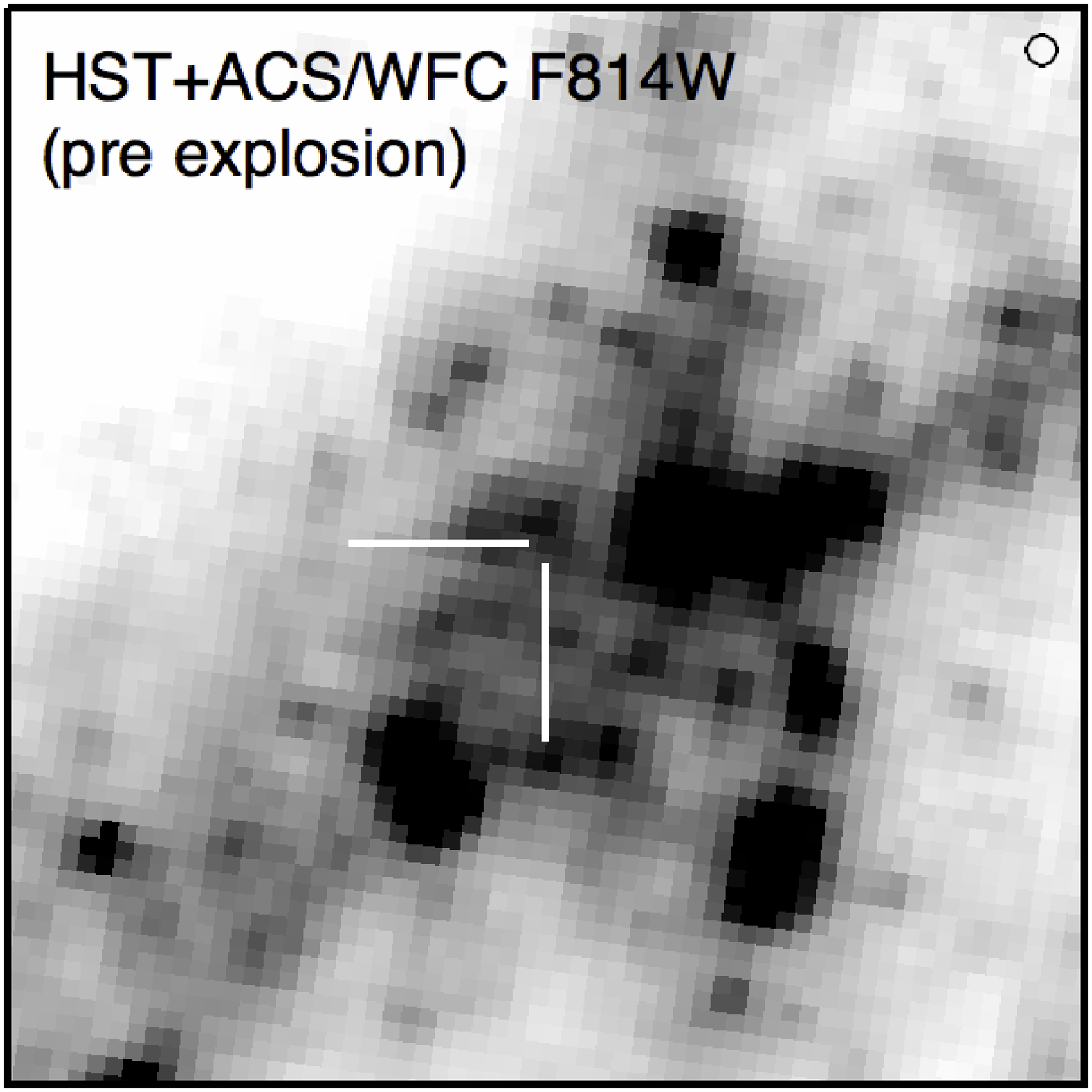}
     \includegraphics[width=0.32\linewidth]{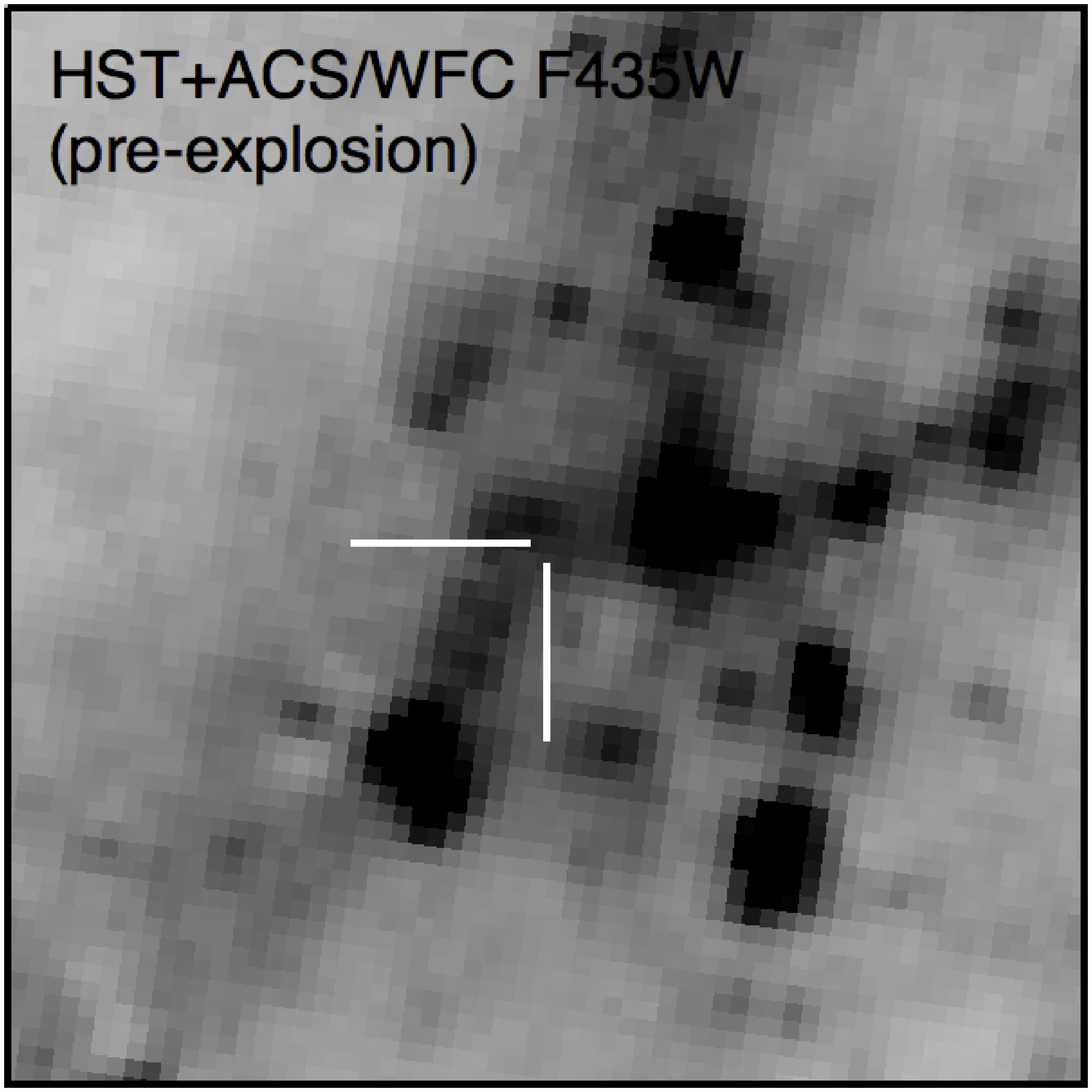}
     \includegraphics[width=0.32\linewidth]{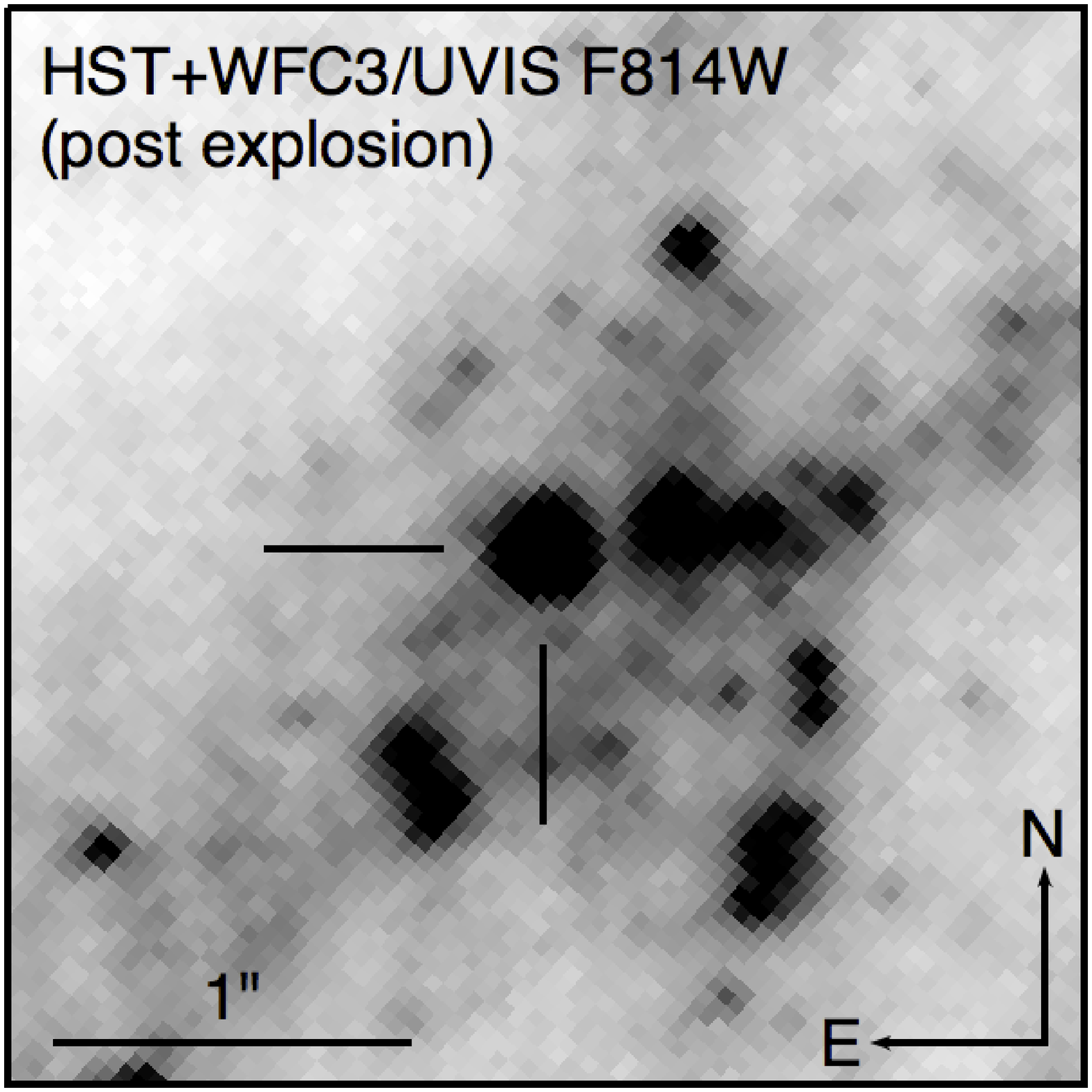}
\caption[]{(a) Pre-explosion \textit{HST}+ACS/WFC \textit{F814W} image of the site of SN~2010O, with the transformed SN position indicated at the centre of the tick marks. The circle in the upper right has a radius five times the positional uncertainty on the pre-explosion image, arising from the transformation and SN position. (b) The pre-explosion \textit{F435W} image. (c) The post-explosion image taken of SN~2010O with \textit{HST}+WFC3/UVIS \textit{F814W}, and which was aligned to the pre-explosion images. The scale and orientation of all three panels are identical and indicated in the panel on the right.}
\label{fig:progenitor_10O}
\end{figure*}

The pipeline drizzled pre-explosion Advanced Camera for Surveys (ACS) Wide Field Channel (WFC) \textit{F814W} image (0.050~arcsec~pixel$^{-1}$) taken on 2006 March 19 (archive file name j9cv38020\_drc, programme GO-10592, PI: A. Evans) was aligned to the post-explosion WFC3/UVIS \textit{F814W} image (0.040~arcsec~pixel$^{-1}$) taken on 2010 June 24 (archive file name ibfz01030\_drz, programme GO-12295, PI: H.~Bond). We identified 21 point-like sources common to both frames and within a 12-arcsec radius of the SN position, and measured their pixel coordinates. The matched pixel coordinates in each image were then used to derive a geometric transformation using {\sc iraf geomap}, with an rms error of only 8~mas (0.16~pixels). The pixel coordinates of the SN were then measured in the post-explosion image using the three centring algorithms (Gaussian, centroid and ofilter) within {\sc iraf phot}; the average was taken as the SN position, and the standard deviation (2~mas) was taken as the uncertainty. Transforming the measured SN coordinates to the pre-explosion image, we do not find a point source coincident with the SN, although we find a source offset by 35~mas (0.69~pixels) as shown in Fig. \ref{fig:progenitor_10O}. This offset is larger than the sum of all uncertainties of the measurements, and at the distance of Arp~299 corresponds to $\sim$8~pc.

The source is clearest in the \textit{F814W} image (approximately corresponding to \textit{I} band), and we measure a magnitude in the VEGAMAG system of $m_{F814W} = 23.115 \pm 0.025$~mag using the {\sc dolphot} package \citep[{\sc dolphot} is a modified version of the {\sc hstphot} package;][]{dolphin00}. Photometry was performed on the \_flc files, which have been corrected for charge-transfer efficiency (CTE) losses, but have not been drizzled. The $\chi^2$ value returned by {\sc dolphot} ($\chi^2 = 4.8$) indicates that the source was not particularly well fit by with a PSF, while the sharpness parameter ($-0.18$) is consistent with a slightly extended source. We do not see an obvious corresponding source in the \textit{F435W}-filter image taken at the same epoch (Fig.~\ref{fig:progenitor_10O}). We also examined the multiple Wide-Field and Planetary Camera 2 (WFPC2) images which covered the site of SN~2010O, but as these were of inferior resolution and depth to the ACS data they have not been considered any further here.

For the distance modulus and line-of-sight extinction towards SN~2010O ($A_{F814W}=1.16$~mag, $\mu=33.26$~mag), the nearby source has an absolute magnitude of $M_{I}\sim-11.3$. This is too bright to be a single star. For instance, if adopting a bolometric correction of 0, this corresponds to a luminosity of $\mathrm{log} L = 6.4$~dex; for comparison the progenitor of SN~2011dh had a luminosity of $\mathrm{log} L = 4.9$~dex \citep{maund11}. It is possible that the source is a cluster with which SN~2010O is physically associated. However, given the distance to Arp~299 and the crowded nature of the region, it is impossible to make this determination with any confidence. We hence conclude that there is no point-like source coincident with SN~2010O, but that this is not surprising given the distance and extinction. Hence we do not set any strong limits on the progenitor luminosity.

Our findings on SN~2010O are in good agreement with the earlier report of \citet{bond11} concluding that the SN is close but not coincident with a very blue stellar cluster. \citet{bond10, bond11} did not study SN~2010P, likely due to the high host galaxy extinction.

\subsection{SN~2010P}

SN~2010P suffers from even higher extinction than SN~2010O ($A_{V}=7$~mag), which together with the distance of Arp 299 makes any search for a progenitor at optical wavelengths futile. We have hence only considered the NIR data in the {\it HST} archive, which consists of a 600-s observation taken with the Near Infrared Camera and Multi-Object Spectrometer (NICMOS) using the NIC2 camera (0.075~arcsec~pixel$^{-1}$) and the \textit{F160W} filter (approximately corresponding to \textit{H} band), taken on 2003 September 1 (data set n8n716010, snapshot programme GO-9726, PI: R. Maiolino). 

\begin{figure*}
\centering
  \includegraphics[width=0.32\linewidth]{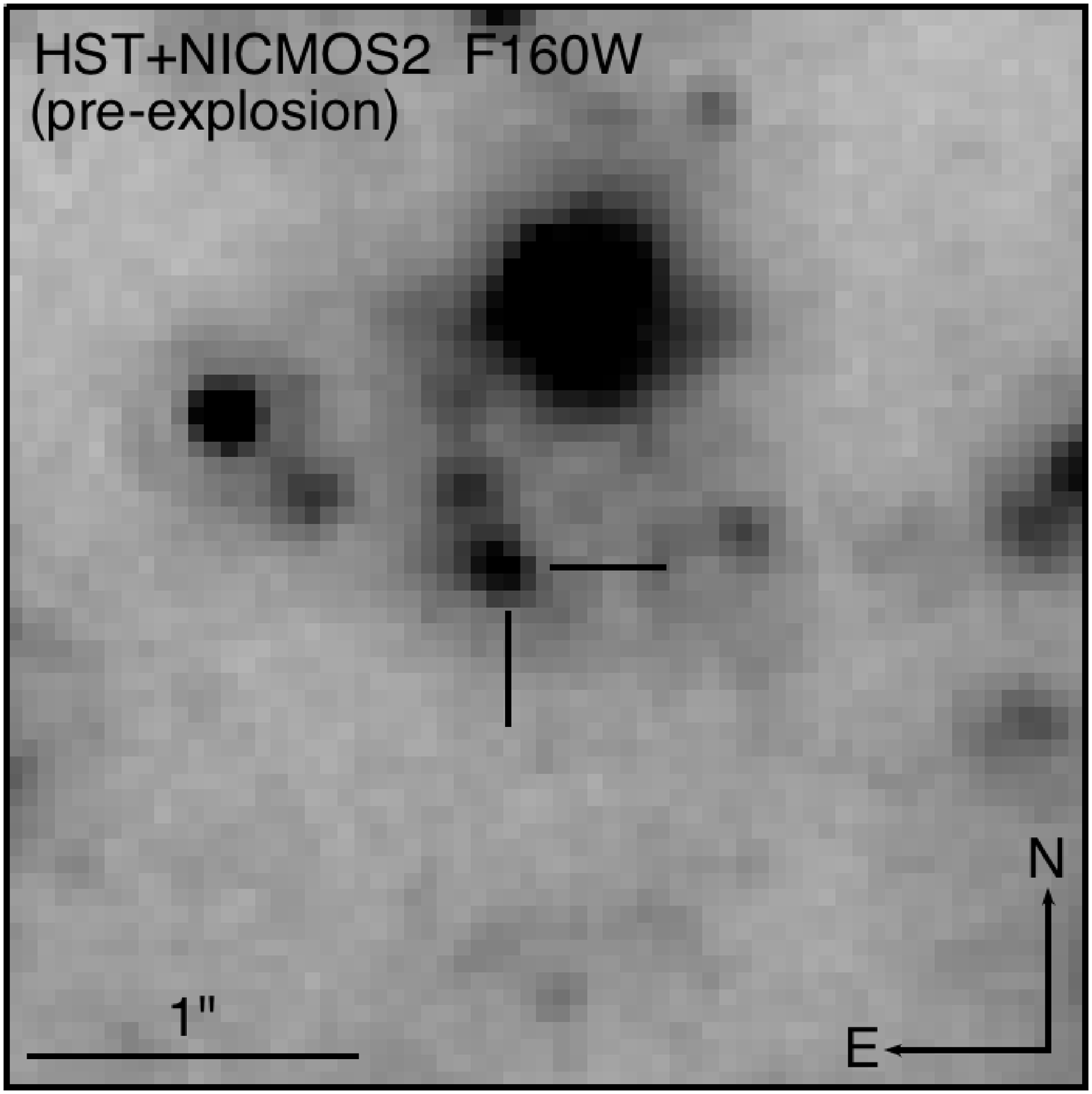}
     \includegraphics[width=0.32\linewidth]{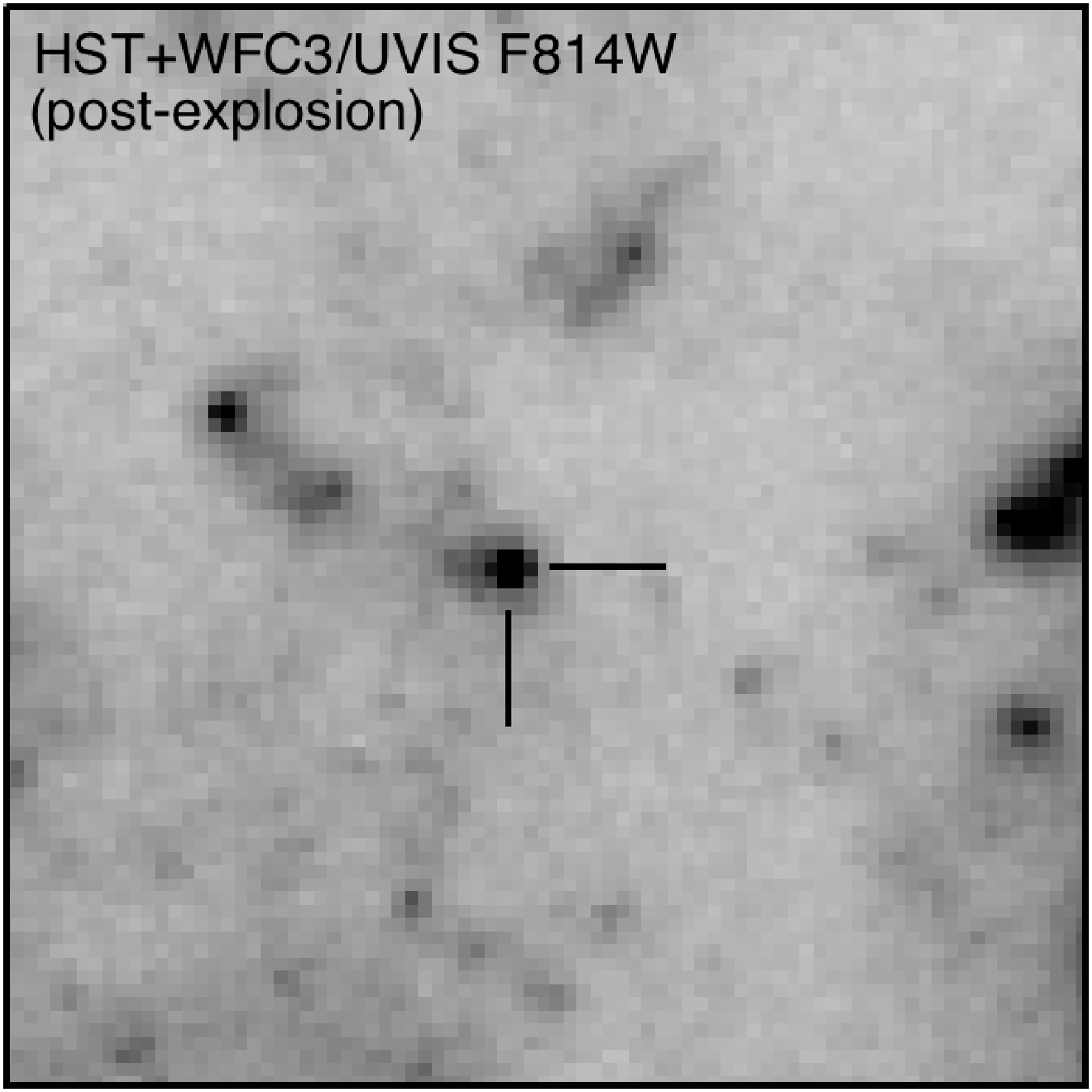}
\caption[]{(a) Pre-explosion \textit{HST}+NICMOS2 \textit{F160W} image of the site of SN~2010P, with the transformed SN position indicated at the centre of the tick marks. (b) The post-explosion \textit{HST}+WFC3/UVIS \textit{F814W} image of SN~2010P which was aligned to the pre-explosion images. Because of the high host galaxy extinction, the SN is not visible in other optical post-explosion images. The scale and orientation of all two panels are identical and indicated in the panel on the left.}
\label{fig:progenitor_10P}
\end{figure*}

Using the same procedure as for SN 2010O, we find a position of SN~2010P in the NIC2 \textit{F160W} image with an uncertainty of 8~mas, as shown in Fig.~\ref{fig:progenitor_10P}. The transformed position is offset by 0.7~pixels (35~mas) from a nearby source. The SN is formally not coincident with this source, although as in the case of SN~2010O, it is possible that the source is a cluster with which the progenitor of SN~2010P was associated. We performed aperture photometry on the source using a small aperture within {\sc iraf}, and using the appropriate photometric calibration from the STScI
webpages,\footnote{\urlwofont{http://www.stsci.edu/hst/nicmos/performance/photometry/postncs\_keywords.html}}
along with aperture corrections derived using {\sc tiny tim},\footnote{\urlwofont{http://tinytim.stsci.edu}} we measure a magnitude for the source of $m_{F160W} = 19.1 \pm 0.2$~mag in the AB system. The absolute magnitude ($M_{F160W}=-15.8$~mag) is too bright to be a single massive star, leaving open the possibility that it is a cluster or complex.

To conclude, both SN~2010O and SN~2010P are close, but not coincident with bright sources. In both cases, these sources are probably clusters or complexes. They are too bright to be single stars, but rather have typical luminosities of super star clusters in such galaxies \citep{randriamanakoto13}. These clusters are typically young with ages of a few Myr, and hence very plausible sites for massive CCSNe. However, since we do not know the exact characteristics of the cluster candidates in question here, nor whether they are with any certainty physically related to SNe~2010O and 2010P, we can infer no information on the progenitors of either SN from these results.

\section{Discussion}

SN~2010O appears to be photometrically and spectroscopically a normal Type Ib SN. A fast decline rate after the light-curve peak suggests that not all the $\gamma$-rays are trapped in the ejecta, which is typical for Type Ib/c SNe. The physical parameters, $E_{\mathrm{k}}(\mathrm{total}) \approx  3 \times 10^{51}$~erg, $M_{\mathrm{ej}}(\mathrm{total}) \approx 2.9$~M$_{\sun}$ and $M_{\mathrm{Ni}}(\mathrm{total}) \approx  0.16$~M$_{\sun}$, inferred for explosion of SN~2010O, would suggest, assuming a single progenitor, a rough zero-age main-sequence mass of $\sim$20$-$25~M$_{\sun}$ \citep[see e.g.][]{mazzali09} which has lost its outer hydrogen envelope via stellar winds. In the case of a binary system, as suggested for SN~2010O also by \citet{nelemans10}, where the outer envelope stripping of the progenitor star would take place via Roche-lobe overflow \citep*{podsiadlowski92}, the progenitor mass can be expected to be smaller. 

Already during post-maximum, some Type IIb and Type Ib SNe do not show significant spectroscopic differences, especially as many Type Ib SNe can in fact show some weak signs of hydrogen in their spectra. The spectrum of SN~2010P, obtained roughly 11 d after the peak, is found to be spectroscopically very similar to such Type IIb events as SNe~2000H, 2008ax and 2011dh. All the above mentioned SNe do not show strong hydrogen features anymore a few days after the peak and relatively low ($<$0.1~M$_{\sun}$) hydrogen masses have been inferred for them \citep{elmhamdi06,chornock11,bersten12}. The \textit{IJHK} light curves of SN~2010P do not show any signs of a double peaked maximum, however, it might be that we lack observations obtained very early and in the UV region for a clear detection of this effect. Nonetheless, the Type IIb SN~1993J, which also shows more prominent hydrogen features at comparable epochs to the spectrum of SN~2010P and has been associated with higher hydrogen mass \citep{elmhamdi06} does show a clear double-peaked light curve. The rise to the first initial light-curve peak is associated with shock heating of the extended low-mass hydrogen envelope followed by a declining light curve and recombination of the ionized hydrogen analogous to Type IIP SNe \citep[e.g.][]{shigeyama94}. Therefore, the weak hydrogen features in the spectrum and the lack of this early light-curve evolution suggest that SN~2010P was not an extreme case of Type IIb SNe with a relatively high mass of the outer hydrogen envelope, but instead had a hydrogen mass of $\lesssim$0.1~M$_{\sun}$ and a more compact progenitor. \citet{claeys11} placed a hydrogen mass lower limit of 0.1~M$_{\sun}$ for Type IIb SNe, which would classify SN~2010P as a Type Ib SN; however, the late-time radio observations (Paper~II) rule out a Type Ib origin of SN~2010P and we consider the object to be a Type IIb event. Interestingly, the very long rise time of hundreds of days to the radio maximum shown by SN~2010P (Paper~II) is empirically expected to occur with Type IIb SNe with much extended ($R \sim 10^{13}$~cm) envelopes with hydrogen mass of $>0.1$~M$_{\sun}$ \citep[][]{chevalier10} rather than compact ($R \sim 10^{11}$~cm) progenitors. However, we note that the case of SN~2010P could be somewhat analogous to SN~2011dh, for which a combination of analysis of high-resolution pre-explosion imaging and hydrodynamical modelling of the light curves has suggested a consistent model with an extended ($R \sim 10^{13}$~cm) progenitor with a modest mass of hydrogen \citep*{maund11,vandyk11,bersten12,benvenuto13}, even when e.g. the early-time light curve has suggested a more compact progenitor \citep[e.g.][]{arcavi11}. However, the rise time of SN~2011dh to the maximum luminosity in radio wavelengths was relatively fast and happened in a few tens of days~\citep{soderberg12}, unlike in the case of SN~2010P (Paper~II). Furthermore, recent results of the Type IIb SN~2011hs \citep{bufano13} have found also a discrepancy between the radio and optical observations, inferring a compact or extended progenitor, respectively. SNe~2010P, 2011dh and 2011hs are therefore examples of transients which suggest that the relation between radio rise time and the progenitor nature of Type IIb SNe is more complex than previously thought. 

The early NIR and optical observations of SN~2010P did not show any obvious signs of ejecta--CSM interaction and the early radio observations yielded non-detections. However, after a long observational gap, at age 1.4~yr from the explosion, the radio luminosity of SN~2010P (Paper~II) reveals that the ejecta is interacting with a CSM, suggesting that the progenitor experienced episodic mass loss at the end of its life cycle. We note that recent modelling results have suggested that some $\sim$20~M$_ {\sun}$ stars could explode as Type IIb SNe and experience variability in their mass-loss rate due to crossing the bistability limit shortly before the core-collapse, resulting in modulations in their radio light curve \citep*[see][and references therein]{moriya13}, consistent with the radio observations of SN~2010P. Furthermore, radio light-curve modulations have been observed in the follow-up of some \textit{compact} Type IIb SNe, such as e.g. SNe~2001ig \citep{ryder04}, 2003bg \citep{soderberg06b}, 2008ax \citep{roming09} and 2011ei \citep{milisavljevic13}, see also the discussions in \citet{kotak06}. However, such observations have not shown as late re-brightening in radio wavelengths as the peak time observed for SN~2010P.

\section{Conclusions}

Based on the analysis of both NIR light curves and optical spectroscopy we find consistent results for the explosion date, SN type, line-of-sight host galaxy extinction and the extinction laws in Arp~299 for both SNe~2010 and 2010P. We find both events to be stripped-envelope SNe and to have exploded by chance very close in time to one another, within only a few days. A simple model for the bolometric light curve of SN~2010O yields a very rough estimate for the SN explosion parameters of $E_{\mathrm{k}} \approx  3 \times 10^{51}$~erg, $M_{\mathrm{ej}} \approx 2.9$~M$_{\sun}$ and $M_{\mathrm{Ni}} \approx  0.16$~M$_{\sun}$, suggesting either a $M_{\mathrm{ZAMS}}$~$\sim$20$-$25~M$_{\sun}$ single progenitor star or possibly a less massive progenitor in a binary system. The reported optical/NIR observations indicate a compact progenitor with a modest amount of hydrogen for SN~2010P, whereas the radio observations (presented in Paper~II) with a long rise time to the radio peak suggest an extended progenitor for the SN. Such a discrepancy indicates that the radio rise time and the progenitor nature relation of Type IIb SNe is more complex than previously suggested. The study of high-resolution pre-explosion images of Arp~299 reveals both SNe to be close to, but not coincident, with extended sources that are likely massive and possibly blended clusters. Because of the distance and high host galaxy extinction, no further constraints on the progenitors are possible based on pre-explosion data. 

Based on both light-curve and spectroscopic comparisons we derive host galaxy extinctions of $A_{V}\approx2$ and $7$~mag for SNe~2010O and 2010P, respectively. The \citet{cardelli89} extinction law is found to be the best choice to describe the host galaxy line-of-sight extinction to SN~2010O in the spiral arm of the A component of Arp~299, whereas in the case of SN~2010P the \citet{calzetti00} extinction law is found to be more consistent with the observations. The latter extinction law is thus found to better describe the dust properties of the IR bright C$\arcmin$ nucleus of Arp~299. Future SN discoveries, with multiwavelength follow-up observations, in the nuclear regions of the major components of Arp~299 could be used to independently map the extinction properties of this prototypical LIRG and to study its SN population. This is a strong motivation for high-resolution IR searches of CCSNe in the high SN rate LIRGs such as Arp~299.

\section*{Acknowledgements}

We thank the anonymous referee for very useful comments. We thank Mattias Ergon for providing the spectrum of SN~2011dh for our comparison. EK acknowledges financial support from the Jenny and Antti Wihuri Foundation. SM and CR-C acknowledge support from the Academy of Finland (Project 8120503). AP is partially supported by the PRIN-INAF 2011 with the project `Transient Universe: from ESO Large to PESSTO'. NE-R acknowledges financial support by the MICINN grant AYA2011-24704/ESP, by the ESF EUROCORES Program EuroGENESIS (MINECO grants EUI2009-04170) and by the European Union Seventh Framework Programme (FP7/2007-2013) under grant agreement no 267251. CR-C acknowledges financial support from the ALMA-CONICYT FUND Project 31100004. AA, RH-I and M-AP-T acknowledge support from the Spanish MINECO Projects AYA2009-13036-CO2-01 and AYA2012-38491-C02-02, cofunded with FEDER funds. This work was partly supported by the European Union FP7 programme through ERC grant number 320360.

Based on observations obtained at the Gemini Observatory, which is operated by the Association of Universities for Research in Astronomy, Inc., under a cooperative agreement with the NSF on behalf of the Gemini partnership: the National Science Foundation (United States), the National Research Council (Canada), CONICYT (Chile), the Australian Research Council (Australia), Minist\'{e}rio da Ci\^{e}ncia, Tecnologia e Inova\c{c}\~{a}o (Brazil) and Ministerio de Ciencia, Tecnolog\'{i}a e Innovaci\'{o}n Productiva (Argentina).

Some of the data presented in this paper were obtained from the Mikulski Archive for Space Telescopes (MAST). STScI is operated by the Association of Universities for Research in Astronomy, Inc., under NASA contract NAS5-26555. Support for MAST for non-\textit{HST} data is provided by the NASA Office of Space Science via grant NNX13AC07G and by other grants and contracts.

This paper is based on observations made with the Nordic Optical Telescope, operated on the island of La Palma jointly by Denmark, Finland, Iceland, Norway and Sweden, in the Spanish Observatorio del Roque de los Muchachos of the Instituto de Astrofisica de Canarias.

The Liverpool Telescope is operated on the island of La Palma by Liverpool John Moores University in the Spanish Observatorio del Roque de los Muchachos of the Instituto de Astrofisica de Canarias with financial support from the UK Science and Technology Facilities Council.

This paper makes use of data obtained from the Isaac Newton Group Archive which is maintained as part of the CASU Astronomical Data Centre at the Institute of Astronomy, Cambridge.

This research has made use of the NASA/IPAC Extragalactic Database (NED) which is operated by the Jet Propulsion Laboratory, California Institute of Technology, under contract with the National Aeronautics and Space Administration.

We have made use of the Weizmann interactive supernova data repository (\urlwofont{www.weizmann.ac.il/astrophysics/wiserep}).


\end{document}